\documentclass{jfm_modified_2021}
\usepackage{graphicx}
\usepackage{amsmath, amssymb}
\usepackage[inactive,tightpage]{preview}

\usepackage{microtype}
\usepackage[normalem]{ulem}
\usepackage{mathtools,bm}
\usepackage{esvect}
\usepackage{enumitem}

\pdfmapfile{+rsfso.map}
\DeclareSymbolFont{rsfso}{U}{rsfso}{m}{n}
\DeclareSymbolFontAlphabet{\mathscr}{rsfso}

\usepackage{pdflscape}
\usepackage{afterpage}
\usepackage{flafter}
\usepackage{rotating}
\usepackage{fancyhdr}
\fancypagestyle{lscape}{%
\fancyhf{} 
\fancyfoot[LE]{}
\fancyfoot[LO] {}
 
}
\usepackage{import}

\usepackage{bm}
\usepackage{microtype}

\usepackage{hyperref}
\usepackage[usenames, dvipsnames]{xcolor}
\hypersetup{colorlinks=true,linkcolor=MidnightBlue,citecolor=MidnightBlue}

\usepackage{cleveref} 
\usepackage{natbib}
\usepackage{array}
\usepackage{booktabs}
\usepackage{multirow}

\usepackage{caption}
\usepackage[position=top]{subcaption}
\usepackage{tabularx}
\usepackage{siunitx}
\newcolumntype{Y}{>{\centering\arraybackslash}X}

\usepackage{tikz}
\usepackage{pgfplots}
\usetikzlibrary{backgrounds, calc, plotmarks, arrows.meta}

\usepackage{grffile}
\pgfplotsset{compat=newest}
\usepgfplotslibrary{patchplots}

\usepackage[outline]{contour} 
\contourlength{0.15em}

\usepackage{lscape}

\newcommand*{\de}{\operatorname{d\!}{}}
\newcommand{\dd}[2]{\frac{\de#1}{\de#2}}

\newcommand{\pd}[2]{\frac{\partial#1}{\partial#2}}

\usepackage{printlen}

\shorttitle{Dimensional analysis of catchments}
\shortauthor{Morawiecki and Trinh}

\title{On the development and analysis of coupled surface-subsurface models of catchments. 
\\[0.2em] \large Part 1. Analysis of dimensions and parameters for UK catchments}

\author{Piotr Morawiecki\corresp{\email{piotr.morawiecki@bath.edu}}
 \and Philippe H. Trinh\corresp{\email{p.trinh@bath.ac.uk}}}

\affiliation{
    Department of Mathematical Sciences, University of Bath, Bath BA2 7AY, UK
}

\pubyear{XXXX}
\volume{YYY}
\pagerange{xxx--xxx}
\date{\today~[Draft]}

\newcommand{\edit}[1]{#1}
\newcommand{\editt}[1]{#1}

\newcommand{\med}[1]{\mathrm{med}\left(#1\right)}

\newcommand{\dt}[1]{\frac{\partial #1}{\partial t}}

\begin{document}

\maketitle



\begin{abstract}
    \noindent The objective of this three-part work is to formulate and rigorously analyse a number of reduced mathematical models that are nevertheless capable of describing the hydrology at the scale of a river basin (\emph{i.e.} catchment). Coupled surface and subsurface flows are considered. 

    In this first part, we identify and analyse the key physical parameters that appear in the governing formulations used within hydrodynamic rainfall-runoff models. Such parameters include those related to catchment dimensions, topography, soil and rock properties, rainfall intensities, Manning's coefficients, and river channel dimensions. Despite the abundance of research that has produced data sets describing properties of specific river basins, there have been few studies that have investigated the ensemble of typical scaling of key physical properties; these estimates are needed to perform a proper dimensional analysis of rainfall-runoff models. Therefore, in this work, we perform an extensive analysis of the parameters; our results form a benchmark and provide guidance to practitioners on the typical parameter sizes and interdependencies. Crucially, the analysis is presented in a fashion that can be reproduced and extended by other researchers, and wherever possible, uses publicly available data sets for catchments in the United Kingdom. 
\end{abstract}


\section{Introduction}
\label{sec:introduction}

\noindent A key problem in hydrology concerns the prediction of water flow into the river. Here, the basic geometry considered is a drainage basin or a catchment site, defined by the area of land from which flow induced by precipitation will reach a specified outlet. As one would expect, the complete physics of this fluid transport problem involves a host of complex multiscale effects related to the precise geological, ecological, urban, and climate-related features of the environment. 

The objective of this work, divided into three parts, is the formulation and rigorous analysis of a number of relatively simple mathematical models that\edit{, following the basic assumptions of coupled surface-subsurface models,} are nevertheless capable of describing the hydrology at the scale of a typical catchment site. In particular, we wish to study the following questions: 
\begin{enumerate}[label={(\roman*)}, leftmargin=*, align = left, labelsep=\parindent, topsep=3pt, itemsep=2pt,itemindent=0pt]
\item What is a relatively simple mathematical model that describes catchment-scale dynamics of subsurface and surface flow? What are the key non-dimensional quantities that govern the physics of such phenomena? What scaling laws can be predicted based on asymptotic analysis of these models?
\item Are these reduced models justifiable given available data of catchments in the United Kingdom? What are the typical parameter values to use in such models?
\end{enumerate}

\noindent The above forms a set of general questions of interest. We may pose much more specific ones of fluid mechanical origin. For instance:
\begin{quotation}\noindent
Given a typical catchment geometry in the United Kingdom with typical length scales, typical terrain slope, typical soil conductivity, and so forth, what are the predicted scaling laws characterising the flow rates into the river during a rainfall of typical intensity and duration?
\end{quotation}

\noindent One challenge is to define the notion of what constitutes a ``typical" parameter; this concerns the second question (ii) we have asked above.

\editt{We wish to study the behaviour of mathematical models and to do so, we require some notion of reasonable parameter values. However, in the context of hydrology, determination of typical parameters is not always possible: real-world catchments are characterised by a wide range of shapes and topography, with hydraulic properties that are highly heterogeneous and may significantly vary across different scales (see discussion by \citealt{clauser1992permeability}). With such high natural variability in catchment properties, it is not possible to formulate one benchmark scenario that reflects the properties of all existing catchments. In this three-part work, our overarching aim to formulate a mathematical model of a catchment that oversimplifies the complexity of real-world catchments, but is characterised by plausible/typical physical parameters.} 

\editt{The task of the present manuscript is to study this issue of parameter values. We wish to obtain the parameter choices in a transparent way via the statistical analysis of real-world data or via references to other works.} In general, we shall estimate the median and interquartile range of each parameter considered for a wide range of UK catchments, and then study their correlations and dependencies. Whenever possible, we use publicly available spatial data. However, some parameters (\emph{e.g.} soil parameters, such as hydraulic conductivity) are hard to measure (or average) at the catchment scale; in these cases, we based our estimates on existing models or individual experimental case studies.

\subsection{On mathematical and computational models}


\noindent \edit{
%
%
We have observed that in standard practices of industrial hydrological research on coupled surface-subsurface flows at the catchment-scale, the emphasis is often on obtaining the exact prediction of flow quantities given available data at specific catchments [see \emph{e.g.} the review by \cite{furman2008modeling} and references therein]. While this approach allows site-specific predictions, there seems to have been less work on the study of the general properties of coupled surface-subsurface flow models. This is a challenging task because of the complexity of real-world catchments, which are characterised by multiple temporal and spatial scales.}

\edit{We are interested in the development of minimal mathematical catchment models that are simple enough to allow the development of universal scaling laws, but at the same time, are complex enough to represent the key physical processes characterising real-world systems. The study of such fundamental properties for simple benchmark models may help in understanding the limitations of such models when applied to the real world catchments---beyond what can be learned from standard numerical approaches.}

We shall present a more thorough literature review of mathematical and computational models in Part 2, but here we review some relevant strands of investigation. First, there are classic references concerning both subsurface flow [by \emph{e.g.} \cite{bear1972dynamics} and \cite{anderson2015applied}] and surface flow [by \emph{e.g.} \cite{chow1959open} and \cite{chow1973hydrodynamic}]. These texts introduce equations of \emph{e.g.} Boussinesq, Richards, Darcy, Saint Venant, and so forth. However, the theories in many of these classical references are not so easily adapted to direct comparison with statistical catchment data in a given location (such as the United Kingdom). Their representation is typically one-dimensional, and in constrained in simplified domains, and without explicit specification of parameters; the analysis is more so qualitative and presents a generalised theory of physics-based modelling. The result, however, is that it is not-at-all obvious how the required scaling laws, raised above in the introduction, can be derived from these isolated theories. 

Generally, modern implementations of the fundamental theory of surface and sub-surface flow do not treat the governing equations in isolation---that is to say, as applied to a simplified mathematical model of a catchment site. Instead, they often take the form of extensive three-dimensional computational models and software (see \emph{e.g.} the introduction by \cite{beven2011rainfall} and reviews by \cite{shaw2015hydrology} and  \cite{bloschl2006rainfall}). In the computational era, this approach has led to the development of codes such as TOP Model~\citep{beven1977towards}, MIKE SHE~\citep{abbott1986introduction1}, HydroGeoSphere~\citep{brunner2012hydrogeosphere}, ParFlow~\citep{kollet2006integrated}, OpenGeoSys~\citep{kolditz2012opengeosys}, and many others. One of our core questions is whether the typical output of a large-scale physics-based model can be explained by a simplified fluid mechanical model.

In addition to physics-based models, many modern references have tended towards statistical or phenomenological modelling. These approaches include predictions of flow rates based on statistical methods, such as multidimensional linear regression~\citep{calver2009comparative}, as well as so-called conceptual rainfall-runoff models~\citep{sitterson2018overview}. A detailed comparative analysis between these three classes of models (statistical, conceptual, and physics-based) is an interesting topic we have highlighted for future work---in some sense, we anticipate that this challenge of inter-model comparison must first begin by agreeing on the minimal mathematical model to consider.

There are challenges to estimating the typical parameters as it is required for further mathematical modelling. In Part 2 of our work, it will be argued that under certain conditions, catchment dynamics can be modelled in terms of simplified geometries where the subsurface and surface flow travels towards the river channel in a transverse direction to the channel flow. Such reduced-dimensional flows will be governed by non-dimensional parameters that involve, for instance, a typical catchment width, say $L_x$, measured in a specific direction. However, given the complex network of streams, rivers, and land topography in any location, it is not clear how $L_x$ should be estimated. Moreover, what is the proper definition of $L_x$ that provides consistency with the underlying assumptions of the model? These and similar questions do not seem to have yet been addressed by existing research.

\subsection{On the development of a reproducible framework for parameter estimation}

\noindent During the course of this work, we have discovered that it is an entirely non-trivial task to seek such \emph{typical} parameters required for mathematical modelling. In many cases, the parameters used by modern software are determined through a black-box calibration of a complex computational model; the details of these procedures are not often published, or their reproduction may be impossible without access to the original codes (see in addition \citealt{hutton2016most}). Consequently, it is important to develop a reproducible framework so that scientific researchers without access to specialised datasets can reproduce our methodology. To this end, we have focused, as much as possible, on the use of publicly available datasets. Furthermore, all numerical algorithms used in this paper are available in a readily applicable form.

For UK catchments, notable examples of datasets include the publicly available National River Flow Archive (NRFA)~\citep{fry2010hydrological}, the 3D Soil Hydraulic Database of Europe created by the European Soil Data Centre~\citep{toth20173d}, the Bedrock Geology Model by the British Geological Society~\citep{waters2016construction} and detailed spatial datasets shared by Ordnance Survey~\citep{lilley2011ordnance}. Boundaries of gauged catchments, for which flow at the outlet is regularly monitored, are defined in the aforementioned NRFA.

In this report, we identify key physical parameters in \cref{sec:fundamentals}, and in \cref{sec:typical_parameters} we use the above datasets to extract typical values of these parameters for all gauged catchments in the United Kingdom.  Mean values of parameters, as well as their correlations and spatial distribution, are investigated in~\cref{sec:statistical_analysis}, followed by discussion in~\cref{sec:discussion}. The goal is to build a foundation for formulation of benchmark scenarios and further dimensional analysis, continued in further parts of our work.

\section{Fundamentals of catchment modelling}
\label{sec:fundamentals}

\noindent Three flows are associated with a general catchment. First, subsurface flow occurs beneath the ground; second, overland flow occurs on what we refer to as the hillslope; third, channel flow occurs within a system of rivers and streams. In this section, we shall review some of the accepted governing equations for these three flows. A more detailed formulation, nondimensionalization, and analysis of simplified catchment models will be presented in Part 2 of our work. Here, our objective is to extract those key dimensional parameters that are expected to be relevant.

\begin{figure}
    \centering
    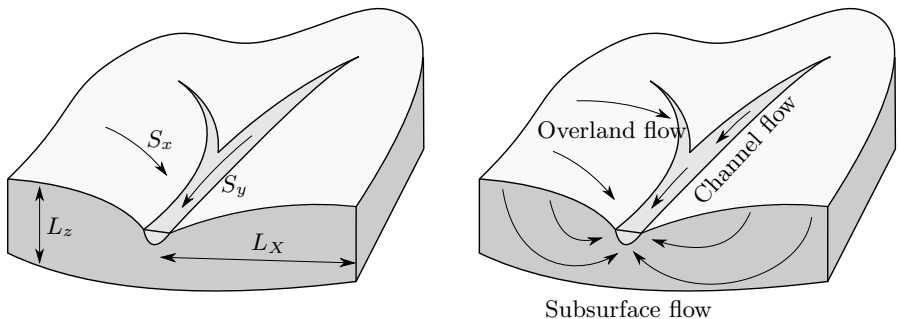
    \caption{On the left, parameters describing catchment shape: hillslope width $L_x$, aquifer depth $L_z$, elevation gradient along the hillslope $S_x$ and along the river $S_y$. On the right, three types of flow are presented; subsurface flow includes both flow through the unsaturated zone and groundwater flow through the saturated zone.}
    \label{fig:catchment_dimensions_and_flows}
\end{figure}

Below, we shall consider a three-dimensional system with a general position vector $\mathbf{x} = (x, y, z)$. The equations presented correspond to general geometries, but for consistency with later mathematical modelling, it will be convenient to associate the $y$ direction as generally oriented along the channel direction; the $x$ direction as generally oriented along the steepest gradient of the typical hillslope; and the $z$ direction as oriented in the vertical direction. Hence, we shall annotate the catchment dimensions with the typical channel length $L_y$, the typical hillslope width $L_x$, and the typical aquifer depth of $L_z$. This geometry is shown in \cref{fig:catchment_dimensions_and_flows}.

\subsection{Subsurface flow}%
\label{sub:subsurface_flow}

\noindent The subsurface flow that governs the pressure head, $h_g(\mathbf{x}, t)$, is commonly modelled using a three-dimensional Richards equation~\citep{dogan2005saturated,weill2009generalized}:
\begin{equation}
    \label{eq:Richards}
    C(h_g)\dt{h_g}=\nabla\cdot \left[K_s K_r(h_g)\nabla(h_g+z)\right].
\end{equation}
Here, $K_s > 0$ is the saturated soil conductivity and $C(h)=\frac{\mathrm{d}\theta(h)}{\mathrm{d}h}$ is the so-called specific moisture capacity. \edit{The pressure head is equal to zero ($h_g=0$) at the  surface of the groundwater table, which separates the fully and partially saturated region of the soil.} We assume that the volumetric water content, $\theta(h)$, and relative hydraulic conductivity, $K_r(h)$, are given by the Mualem-van Genuchten (MvG) model~\citep{van1980closed}: 

\begin{equation}
    \theta(h) =
    \begin{cases} 
        \theta_r+\frac{\theta_s-\theta_r}{\left(1+\left(\alpha_\mathrm{MvG}h\right)^n\right)^m} & h<0 \\
        \theta_s & h\ge0
    \end{cases},
    \label{eq:GMtheta}
\end{equation}

\begin{equation}
    K_r(h) =
    \begin{cases}
        \frac{\left(1-\left(\alpha_\mathrm{MvG}h\right)^{n-1}\left(1+\left(\alpha_\mathrm{MvG}h\right)^n\right)^{-m}\right)^2}{\left(1+\left(\alpha_\mathrm{MvG}h\right)^n\right)^{m/2}} & h<0 \\
        1 & h\ge0
    \end{cases}.
    \label{eq:GMk}
\end{equation}

In essence, the MvG model describes the key hydraulic properties of the soil, such as hydraulic conductivity and saturation, as nonlinear functions of the pressure head, $h$, as well as other parameters $\theta_r$, $\theta_s$, $\alpha_\mathrm{MvG}$, $n$, and $m=1-\frac{1}{n}$. The residual water content, $\theta_r$, and saturated water content, $\theta_s$, represent the lowest and the highest water content, respectively. The parameter $\alpha_\mathrm{MvG}$ $\left[\mathrm{m}^{-1}\right]$ is a scaling factor for pressure head $h$ [m]. Finally, the coefficient, $n$, characterises the distribution of pore sizes. \edit{Since the MvG model parameters describe the soil/rock properties at a given location, in general, they are functions of the spatial coordinates $(x,y,z)$.}

\subsection{Overland flow}%
\label{sub:surface_flow}

\noindent If rainfall exceeds the infiltration capacity, water can accumulate on the surface and form an overland flow. Typically, following \emph{e.g.}~\cite{chow1973hydrodynamic}, \cite{tayfur1994spatially} and~\cite{liu2004two}, this flow is described by the two-dimensional Saint Venant equations, which govern the overland water height, $z = h_s(x, y, t)$. The first equation is the continuity equation for the overland flow, written as
\begin{equation}
    \label{eq:St_Venant_overland}
    \dt{h_s}=\nabla\cdot\mathbf{q}+R-I-E,
\end{equation}
where the source term includes the precipitation rate $R = R(x,y,t)$, the infiltration rate $I = I(x,y,t)$, and the evapotranspiration rate $E = E(x,y,t)$. \edit{The flow $\mathbf{q}$ can be expressed as $\mathbf{q}=h_s\mathbf{u}_s$, where $\mathbf{u}_s = \mathbf{u}_s(x,y,t)$ is the mean flow speed.}

\edit{Therefore, \eqref{eq:St_Venant_overland} is a continuity equation with two unknowns, $h_s$ and $u_s$ (or $q_s$). The second equation required to close this system is provided by momentum conservation. In the general form of the Saint Venant equations, the following equation is used:
\begin{equation}
    \label{eq:SV_momentum_eq}
    \dt{\mathbf{u}_s}+\mathbf{u}_s\nabla{}\cdot \mathbf{u}_s+g\nabla h+g\left(\mathbf{S_f}-\mathbf{S_0}\right)=0,
\end{equation}
where $g$ is the gravitational acceleration, $\mathbf{S_0}$ is the elevation gradient (bed slope), and $\mathbf{S_f}$ is the friction slope, \emph{i.e.} the rate at which energy is lost along the x- and y-directions. One can model $\mathbf{S_f}$ by specifying the shear stress between the overland flow and the surface. However, in the computational hydrological  models, rather than solving eqn \eqref{eq:SV_momentum_eq} directly, the flux, $\mathbf{q}=h\mathbf{u}$ is often obtained using an empirical relationship known as \emph{Manning's equation} [see~\citet[chap.~14.3]{shaw2015hydrology} for more details]. This equation, originally formulated to describe a turbulent channel flow, is also commonly applied to the two-dimensional turbulent overland flow over a rough terrain}. In the vector form, this equation can be written as
\begin{equation}
    \label{eq:Manning}
    \mathbf{q}(\mathbf{x}, h_s) = \frac{1}{n_s}\frac{\mathbf{S_f}}{\sqrt{\|S_f\|}}h_s^{5/3},
\end{equation}
\noindent where  $n_s$ is an empirically determined value known as Manning's coefficient and describes the land surface roughness. Following \eqref{eq:SV_momentum_eq}, the friction slope $\mathrm{S_f}$ is expressed as 
\begin{equation}
    \label{eq:friction_slope}
    \mathbf{S_f}=\mathbf{S_0}-\nabla{h_s}-\frac{1}{2g}\nabla{v^2}-\frac{1}{g}\dt{\mathbf{v}}
\end{equation}
Often in practice, the last two terms of \eqref{eq:friction_slope} can be neglected; this forms the so-called \textit{diffusion approximation}. An additional common simplification is the {\itshape kinematic approximation}, in which the second term in \eqref{eq:friction_slope} is also neglected, giving:
\begin{equation} \label{eq:friction_slope_kin}
    \mathbf{S_f}=\mathbf{S_0}.
\end{equation}
\edit{Note that in general, the gradient varies spatially, therefore $\mathbf{S_0}$ is a function of spatial coordinates $(x,y)$. Similarly, $n_s$ varies not only spatially, but also depends on time through seasonal variations in vegetation \citep{song2017seasonality}.}

\edit{There is an ongoing discussion whether using the above two-dimensional model \eqref{eq:St_Venant_overland}-\eqref{eq:friction_slope} is an appropriate model for the overland flow, since the actual overland flow reaches the channel through a series of channels (natural or artificial ones), rather than being a uniform sheet of surface water $h_s(x,y)$ (see an overview by~\citealt{leibowitz2018connectivity}). Nevertheless, we use this formulation as the current standard used in computational physical catchment models.}

\subsection{Channel flow}%
\label{sub:channel_flow}

\noindent Finally, consider the channel flow illustrated in \cref{fig:catchment_dimensions_and_flows}. We ignore the flow dynamics in the transverse directions of the channel and consider a channel located along $(x(s), y(s))$, for a parameterisation parameter, $s$, defined as the distance from the catchment outlet measured along the channel. Then, the channel water height is described by the Saint Venant equation applied to the height $z = h_c(s, t)$. There are three main differences between the channel flow equations and the Saint Venant equations used in the two-dimensional formulation of overland flow in \cref{sub:surface_flow}: (i) the direction of flow is given by the direction of the channel; (ii) precipitation adds a negligible contribution to the channel flow -- instead, the channel flow is primarily affected by both the surface flow passing through the channel perimeter and the overland flow passing over the river banks; and (iii) the roughness is introduced over the entire perimeter of the channel (e.g., in the case of a rectangular channel, its bottom and the submerged part of its walls).

Mathematically, these assumptions lead to the following governing equation  \citep{vieira1983conditions, chaudhry2007open}:
\begin{equation}
    \label{eq:St_Venant_channel}
    \edit{\dt{A} = } \, w\dt{h_c} = q_\mathrm{in} - \pd{q}{s},
\end{equation}
\edit{where $A = A(s, h_c)$ is the channel cross-section, $w = w(h_c, s) = \dd{A}{h_c}$} is the channel width (constant in the case of a rectangular channel), and $q=q(s,t)$ is the mean flow in the channel. The source term, $q_\mathrm{in}$, is given by considering the total surface and subsurface inflow into the river; hence $q_\mathrm{in}$ is linked to the boundary values of the solutions of Richards~\eqref{eq:Richards} and 2D Saint Venant equations~\eqref{eq:St_Venant_overland}. As for the overland equations, the flux, $q = \mathrm{area} \times \mathrm{velocity} = Av$, rather than being solved using the full momentum equation, is again assumed to be given by the empirical Manning's law, which in the case of a channel takes the form:
\begin{equation}
    \label{eq:Manning_channel}
    q = A\frac{\sqrt{S_f}}{n_c}\left(\frac{A}{P}\right)^{2/3},
\end{equation}
\noindent where $P = P(s, h_c)$ is the channel wetted perimeter, and $n_c$ is Manning's coefficient dependent on the banks and channel bed roughness. The quantity $S_f$ is the friction slope, as defined by~\eqref{eq:friction_slope} (or its simplified forms), with $S_0$ representing the elevation gradient along the river, denoted here as $S_y$. In the case of a rectangular channel, $A=wh_c$ and $P=w+2h_c$. Additionally, we denote the surface water height ($h_c$) at the outlet as $d$. As before, the kinematic approximation \eqref{eq:friction_slope_kin} is often assumed.

\edit{We use the above Manning's law since it is a standard approach in computational hydrological  models. However, there may be some scenarios in which this assumption may not be a valid approach. In supercritical flows (such as in the case of flash floods), the flow may rapidly vary along the $y$ direction, and other approaches should be considered instead (see \emph{e.g.} \citealt{mujumdar2001flood}).}

This completes our formulation of the system of three coupled PDEs used to describe the key water flow components at the catchment scale, namely Richards equation~\eqref{eq:Richards}, the 2D Saint Venant equations for overland flow~\eqref{eq:St_Venant_overland}, and the 1D Saint Venant equation for channel flow~\eqref{eq:St_Venant_overland}.

\subsection{Boundary and initial conditions}%
\label{sub:boundary_and_initial_conditions}

\noindent The solution of the governing partial differential equations \eqref{eq:Richards}, \eqref{eq:St_Venant_overland}, \eqref{eq:St_Venant_channel} for the respective subsurface $h_g$, surface $h_s$, and channel $h_c$ flows must be accompanied by appropriate boundary and initial conditions. For instance, boundary conditions are required to specify the mass exchange between flow components, and these conditions may introduce additional catchment-dependent parameters such as the channel depth and the particular geometry at the river outlet. Computational models also require the specification of initial conditions, which are generally unknown. Typically, the simulations can be run for a burn-in period to allow the results to be independent of the initial conditions. Example formulations of boundary and initial conditions will be studied in Part 2 of our work, where we focus on the mathematical and numerical analysis of model catchments.

\begin{table}
    \centering

    \begin{tabular}{rccll}
        \textsc{parameter} & \textsc{symbol} & \textsc{dimension} \\[0.5em]
        channel length & $L_x$ & $L$ & \multirow{3}{*}{\hspace{-1em}$\left.\begin{array}{l} \\ \\ \\ \end{array}\right\rbrace$} & \multirow{3}{3.3cm}{catchment dimensions (sections \ref{sec:catchment_size} and \ref{sec:soil_rock_propeties})} \\
        hillslope width & $L_y$ & $L$ & \\
        aquifer thickness & $L_z$ & $L$ & \\[0.5em]
        
        slope along the hillslope & $S_x$ & $-$ & \multirow{2}{*}{\hspace{-1em}$\left.\begin{array}{l} \\ \\ \end{array}\right\rbrace$} & \multirow{2}{3.3cm}{catchment topography (\cref{sec:catchment_topography})} \\
        slope along the river & $S_y$ & $-$ & \\[0.5em]
        
        saturated soil conductivity & $K_s$ & $LT^{-1}$ & \multirow{5}{*}{\hspace{-1em}$\left.\begin{array}{l} \\ \\ \\ \\ \\ \end{array}\right\rbrace$} & \multirow{5}{3.3cm}{soil and rock properties (\cref{sec:soil_rock_propeties})} \\
        MvG model $\alpha$ parameter & $\alpha_\mathrm{MvG}$ & $L^{-1}$ & \\
        residual water content & $\theta_r$ & $-$ & \\
        saturated water content & $\theta_s$ & $-$ & \\
        measure of the pore-size distribution & $n$ & $-$ & \\[0.5em]
        
        precipitation rate & $R$ & $LT^{-1}$ & \multirow{4}{*}{\hspace{-1em}$\left.\begin{array}{l} \\ \\ \\ \\ \end{array}\right\rbrace$} & \multirow{4}{3.3cm}{water balance terms (\cref{sec:water_balance})} \\
        runoff per unit area & $Q$ & $LT^{-1}$ & \\
        evapotranspiration rate & $E$ & $LT^{-1}$ & \\
        peak annual precipitation & RMED & $LT^{-1}$ & \\[0.5em]
        
        Manning's coefficient for the land surface & $n_s$ & $L^{-1/3}T$ & \multirow{2}{*}{\hspace{-1em}$\left.\begin{array}{l} \\ \\ \end{array}\right\rbrace$} & \multirow{2}{3.3cm}{Manning's coefficients (\cref{sec:mannings_coefficient})} \\
        Manning's coefficient for the channel & $n_c$ & $L^{-1/3}T$ \\[0.5em]
        
        channel's width & $w$ & $L$ & \multirow{2}{*}{\hspace{-1em}$\left.\begin{array}{l} \\ \\ \end{array}\right\rbrace$} & \multirow{2}{3.3cm}{channel dimensions (\cref{sec:channel_dimensions})} \\
        channel's depth & $d$ & $L$ & 
        
    \end{tabular}
    \caption{List of parameters appearing in the formulation of the integrated catchment model, and where they are discussed in this work. \edit{As discussed in the text, these parameters vary spatially, and in some cases also temporally.}}
    \label{tab:list_of_parameters}
\end{table}

\subsection{Typical values of model parameters reported in the literature}

\noindent All parameters appearing in the equations are hence summarised in \cref{tab:list_of_parameters}. Firstly, let us provide a brief review of the typical values of these parameters, known from the literature:

\begin{enumerate}[label={(\roman*)}, leftmargin=*, align = left, labelsep=\parindent, topsep=3pt, itemsep=2pt,itemindent=0pt]

\item The range of saturated soil conductivity $K_s$ can vary from $10^0$ to $10^{-3}~\mathrm{ms}^{-1}$ for very productive aquifers (for well-sorted sand and gravel, and highly fractured rocks) to below $10^{-9}~\mathrm{ms}^{-1}$ for impervious rocks~\citep{bear1972dynamics}.

\item According to \cite{chow1959open}, Manning's roughness coefficient for channels, $n_c$, can vary from $0.01~\mathrm{sm}^{-1/3}$ for artificial (\emph{e.g.} cement) channels to over $0.1~\mathrm{sm}^{-1/3}$ for channels with dense vegetation. Its value for flood plains, $n_s$, can vary from $0.03~\mathrm{sm}^{-1/3}$ for pastures and cultivated areas without crops to $0.1~\mathrm{sm}^{-1/3}$ in densely forested areas. \edit{In addition, the value of $n_c$ for a specific area can also vary with time due to seasonal variation in vegetation.}

\item The evapotranspiration rate, $E$, in computational physics-based catchment models is usually estimated using models such as the evapotranspiration model by \cite{kristensen1975model} and the Two-Layer UZ/ET model by \cite{yan1994simulation}, both of which are for example used in the MIKE SHE integrated model. However, the formulation of these models is beyond the scope of this report. \cite{cole1991reliable} reports the mean monthly precipitation and evaporation values for different regions of the United Kingdom. Precipitation $R$ highly depends on the UK region, varying from $645.5$~mm/year in Central and East England to $1601.9$~mm/year in Northwest and North Scotland, with the highest precipitation levels observed between October and January. According to \cite{faulkner1998mapping}, the highest daily precipitation measured throughout the year varies from $25$~mm (during a single day) in the East of England to over $80$~mm in some sites in mountainous regions of Wales and Scotland. The evapotranspiration rate $E$ is similar for all regions, but highly varies in time, from $6-12$~mm/month in January to $63-78$~mm/month in July.

\item Apart from specifying the precipitation, one also needs to specify catchment geometry -- its size, terrain topography, and aquifer depth. The publications on integrated catchment models mostly focus on one or a few real-world catchments. There are also works aiming to understand the characteristic properties of catchments and channel networks. Many catchment characteristics were described by \cite{horton1932drainage}, including drainage density, $D_D$, defined as the ratio of total stream length, $L$, and catchment area, $A$ (the authors found typical values for investigated catchments to be $A = 0.64$--$1.367~\mathrm{km}^{-1}$), average distance between streams ($0.73$--$1.56$~km), average overland flow distance ($0.38$--$0.80$~km), slope along the streams ($0.014$--$0.038$), and slope along the land ($0.072$--$0.177$). The cited values were estimated manually based on topographic maps. Together with the development of computing power and the collection of Digital Terrain Models (DTM), the methods for estimation of the above quantities were automated~\citep{tarboton2001advances}.

\item \cite{grieve2016long} compared three different estimation methods for the hillslope width, $L_x$, by estimating its distribution for a number of catchments in the USA. The first estimate was obtained by dividing the catchment area, $A$, by twice the total stream length, $2L$, which is equivalent to the drainage density $(2D_D)^{-1}$. Note that the factor of $2$ accounts for each stream having a hillslope on either side. The second method used a DTM model to find streamlines following the direction of the steepest descent. The lengths of the streamlines are interpreted as a hillslope width. The last method applied by the authors used a slope–area plot to separate areas dominated by channels from hillslopes. The hillslope width, $L_x$, defined using the streamline (flow routing) method is higher than estimates using the drainage density method, however both have a similar order of magnitude ranging from $30$ to $130$~m.

\item A systematic study of the typical parameter ranges and their effect on model predictions was done by \cite{doummar2012simulation}. The authors used the MIKE SHE software for a karst system (the Gallusquelle spring in the Southwest Germany). \edit{The model parameters were calibrated to minimise the root-mean-square error of the daily observed discharge, and the relative importance of each parameter was numerically investigated using sensitivity analysis.} The most significant parameters include the hydraulic conductivity, the moisture content of the unsaturated rock matrix and van Genuchten parameters.

\end{enumerate}

\section{Data sources and processing methods}
\label{sec:typical_parameters}

\noindent As we have noted in \cref{sec:introduction}, there have been a lack of studies that have attempted to collate and analyse the collection of dimensional parameters listed in \cref{tab:list_of_parameters} as a whole, particularly with the intention of further mathematical modelling. Our focus in this section is to describe the data processing techniques that we have used, in order to extract typical values of the physical parameters characterising UK catchments.
Crucially, we have made these tools available publicly in a GitHub repository via \cite{github1}. 
The data sources are primarily from openly-sourced data on catchments in the United Kingdom, but we expect that the methodology can be applied similarly to data from other locations.

\subsection{Catchment dimensions (\texorpdfstring{$L_x$}{}, \texorpdfstring{$L_y$}{})}
\label{sec:catchment_size}

\noindent \edit{Capturing catchment dimensions and topography is challenging, as real-world river networks have a fractal-like structure \citep{rodriguez2001fractal}. An ambiguous quantity concerns the estimation of the river length characterising the catchment length, $L_y$, since there are multiple ways in which it can be defined, and furthermore, its value depends on the precision of the dataset used for estimation. Similarly, the estimate of the catchment/hillslope length, $L_x$, is challenging on account of the high spatial variation and ambiguous definition. The aquifer thickness, $L_z$, which depends on the properties of the soil, will be discussed in \cref{sec:soil_rock_propeties}.}

\begin{figure}
    \centering
    \includegraphics[width=\linewidth]{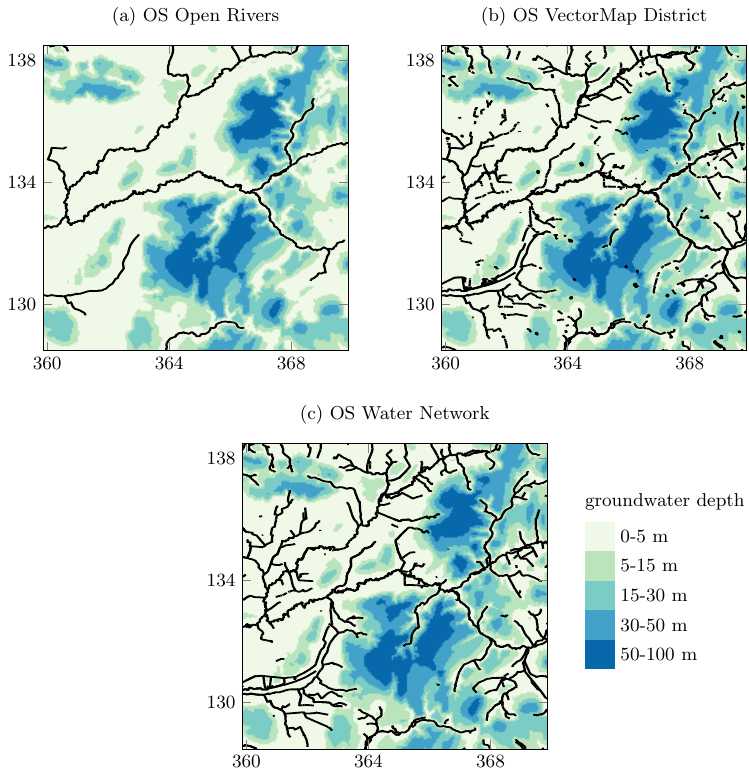}
    \caption{Comparison of three spatial datasets of river locations (thick black lines). The (b) OS VectorMap District and (c) OS Water Network datasets offer a similar level of accuracy. The background colours represent groundwater depth from the BGS Groundwater Levels dataset in meters, indicating the areas of missing streams in the (a) OS Open Rivers dataset, where groundwater reaches the surface in a characteristic finger-like pattern. The northing and easting on the axes are expressed in kilometers.}
    \label{fig:river_datasets_comparison}
\end{figure}

In our work, we use three Ordnance Survey datasets, providing different data formats and levels of detail:

\begin{enumerate}[label={(\roman*)},leftmargin=*, align = left, labelsep=\parindent, topsep=3pt, itemsep=2pt,itemindent=0pt]
\item OS Open Rivers only consists of major rivers represented as spatial lines.
\item OS VectorMap District includes all surface water bodies -- wide rivers and standing bodies of water (\emph{e.g.} lakes and ponds) are represented using spatial polygons, while narrow streams, artificial channels, and ditches are represented using spatial lines.
\item OS Water Network includes all rivers/streams forming the drainage network, but data can be downloaded only for small user-specified regions, not for the entire UK like in the case of the other two datasets.
\end{enumerate}
Fig. \ref{fig:river_datasets_comparison} demonstrates a comparison of the typical information provided by the three types of datasets.

\subsubsection{Estimating the catchment length, $L_y$} \label{sec:Ly}

\begin{figure}
    \centering
    \includegraphics[width=\linewidth]{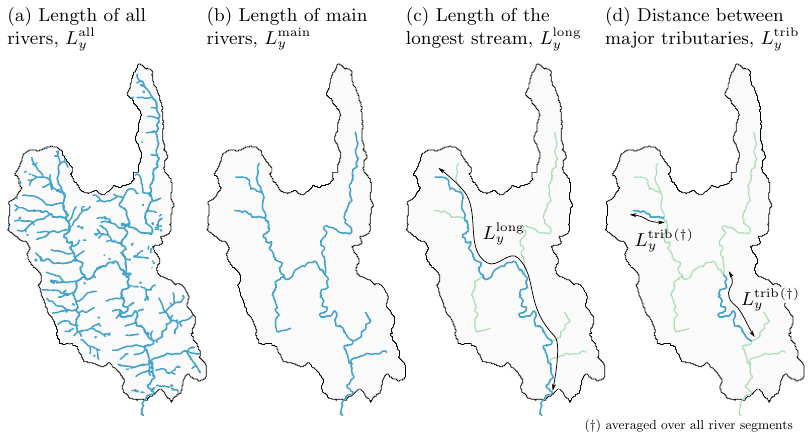}
    \caption{\edit{Illustration of the different length scales characterising the drainage network, presented for the river Frome catchment, with the outlet located at Bishop's Frome. \editt{Each $L_y$ estimate is equal to the total length of all highlighted streams. Note this total length gradually decreases from (a) to (d), therefore $L_y^\text{all}\leq L_y^\text{main} \leq L_y^\text{long} \leq L_y^\text{trib}$.}}}
    \label{fig:length_estimators}
\end{figure}

\noindent In order to estimate the characteristic scale of the catchment length ($L_y$), we propose the following four measures, \edit{graphically represented in \cref{fig:length_estimators}}. Note that depending on their size, the OS Open Rivers dataset represent rivers in the form of polygons (we refer to them as major rivers) or lines (minor rivers). 
\begin{enumerate}[label={(\roman*)}, leftmargin=*, align = left, labelsep=\parindent, topsep=3pt, itemsep=2pt,itemindent=0pt]
    \item Length of main rivers ($L_y^\text{main}$) -- the total length of rivers as defined in the OS Open Rivers dataset; we can use this measure to estimate the total flow into the drainage network.
    \item Length of all rivers ($L_y^\text{all}$) -- the total length of major rivers and minor streams as defined in the OS VectorMap District dataset; it serves the same function as the previous measure, but at a higher spatial resolution; we also use this measure later as one way of measuring the catchment/hillslope width.
    \item Length of the longest river ($L_y^\text{long}$) -- the length of the longest river measured from the spring to the catchment's outlet, extracted from the OS Open Rivers dataset (this measure is approximately the same regardless of whether the minor rivers are included or not); it may be used to investigate the characteristic length of the channel when studying the channel flow, given by~\eqref{eq:St_Venant_channel}. 
    \item Distance between river tributaries ($L_y^\text{trib}$) -- the average distance between river tributaries estimated from the OS Open Rivers dataset, \edit{including the first-order streams (see \cref{fig:length_estimators}d)}; it is the only intensive quantity in this list (\emph{i.e.} it does not scale with the catchment size) and therefore can be used as the characteristic length of a river in which hillslope flow is not disturbed by the presence of tributaries.
\end{enumerate}

\noindent Calculating each of the above measures poses some challenges. Firstly, the river banks and other natural boundaries have a fractal structure, which means that their length can be very sensitive to the chosen spatial resolution. Here, we have used the data in the original resolution to maintain high accuracy in the case of estimates (i)--(iii). In the case of estimate (iv), we are interested in the typical lengths of hillslopes rather than river streams. In this case, the length of the former should not be affected by small-scale local meanders. Therefore, in case (iv), before measuring the river length, we simplify the geometry using the Douglas-Peucker algorithm \citep{douglas1973algorithms} with tolerances of $1000$~m. This algorithm effectively smooths river meanders shorter than the chosen tolerance \edit{and follows a similar methodology as in \cite{stuetzle2009evaluating}}.

Secondly, when using the VectorMap District dataset for calculating the total river/stream length, data corresponding to standing bodies of water (lakes and ponds) should not be taken into account. This cannot be easily implemented since such standing bodies are represented in the same way as for wide rivers within the dataset. Nevertheless, since lakes have much shorter total boundary length than rivers for the majority of UK regions, including such data does not significantly impact the estimates. Therefore, we include all surface water bodies when estimating (ii), which we obtain by adding the sum of the lengths of all spatial lines and the sum of the perimeters of all spatial polygons divided by two (since each of the two banks is counted separately).

\edit{For each UK catchment specified in the National River Flow Archive, we calculated the total length of all major rivers ($L_y^\text{main}$), and all major and minor rivers ($L_y^\text{all}$), as well as the length of the longest stream ($L_y^\text{long}$), and the mean distance between major tributaries ($L_y^\text{trib}$). Note that from their definition \editt{for each catchment we have}, $L_y^\text{all}\leq L_y^\text{main} \leq L_y^\text{long} \leq L_y^\text{trib}$, (see \cref{fig:length_estimators}). \editt{Their median values taken over all considered catchments are: $\med{L_y^\text{main}}=71.6$~km, $\med{L_y^\text{all}}=130$~km, $\med{L_y^\text{long}}=22.8$~km, and $\med{L_y^\text{trib}}=945$~m.}}

\subsubsection{Estimating the catchment width, $L_x$}
\label{sec:estimating_Lx}

\noindent Estimating the characteristic width of the catchment/hillslope, $L_x$, is also challenging due to the high spatial variation and the ambiguous definition of this distance. We use the following two alternative measures for $L_x$:
\begin{enumerate}[label={(\roman*)}, leftmargin=*, align = left, labelsep=\parindent, topsep=3pt, itemsep=2pt,itemindent=0pt]
\item the ratio of total stream length to catchment area ($L_x^\text{area}$); and
\item the length of overland streamlines reaching a given river or river network ($L_x^\text{stream}$).
\end{enumerate}

In (i), the first estimate of $L_x^\text{area}$ is suggested by \emph{e.g.} \cite{rodriguez2001fractal}. Here, we express the catchment area as $A=2L_x^\text{area}L_y^\text{all}$, where the factor of two represents the hillslope on the left- and right-hand bank. This provides an estimate of $L_x^\text{area}$ given the area, $A$, and $L_y^\text{all}$ (see \cref{sec:Ly}). We use $L_y^\text{all}$ since it approximates all streams in the catchment. This estimate is easy to compute, but is sensitive to river boundary roughness. For example, in the case of meandering streams, $L_x^\text{area}$ computed in this fashion can be very large compared to a catchment of identical proportions, but with smooth river banks. 

In (ii), the second estimate of $L_x^\text{stream}$ in (ii) is obtained by estimating the average stream-flow distance between points in the catchment to the nearest stream. Here, we use a sampling method in which a large number of spatial points are distributed over a given area. For each point, we find a streamline from the given point to the stream, following the steepest descent path (which approximates the direction of the overland flow). We use the Digital Terrain Model (DTM) from the OS Terrain 50 dataset is used in order to find the steepest descent directions, while the OS VectorMap dataset is used to determine when such a streamline reaches the river or another surface water body. 
 
The implementation details of (ii) are described in \cref{app:extracting_hillslope_width_gradient}. Since even in catchments with a constant hillslope width $L_x^\text{stream}$, the streamline lengths are distributed uniformly between $0$ and $L_x^\text{stream}$, the mean (and median) distance $\langle\mathrm{dist}\rangle$ is equal to $\langle\mathrm{dist}\rangle=L_x^\text{stream}/2$. Therefore, we estimate the catchment width as $L_x^\text{stream}=2\langle\mathrm{dist}\rangle$. This approach, unlike the previous one, is not sensitive to the roughness of the river boundary; however, it is much more computationally demanding. Using this approach, we found distances to the nearest stream for over 57,000,000 points uniformly distributed over the entire United Kingdom (excluding Northern Ireland). By finding the median value of this distance for each investigated catchment, we estimated the catchment width. 

Summarising the above measures for the UK: using estimate (i), we estimate a median value of $L_x^\text{area} \approx 683$~m when using the ratio of catchment area to the total stream length. Interestingly, this average is close to the median value of estimate (ii), $L_x^\text{stream} \approx 616$~m, which uses the stream-flow definition. Further discussion appears in \cref{app:spatial_distribution} where we remark that $L_x$ significantly varies across different regions of the UK. Denser drainage networks can be found in areas with higher precipitation rates and significant overland flow.

\subsection{Catchment topography (\texorpdfstring{$S_x$, $S_y$}{})}

\label{sec:catchment_topography}

\noindent The gradient of the terrain is important in order to estimate the size of the surface flow, as given by Manning's law~\eqref{eq:Manning}. In order to estimate the typical values of the elevation gradient perpendicular to the river ($S_x$) and along the river ($S_y$), we use the Digital Terrain Model (DTM) from the OS Terrain 50 dataset. 

\subsubsection{Estimating the gradient perpendicular to the river, $S_x$}
\label{sec:estimating_Sx}

\noindent One way to estimate $S_x$ is by plotting the valley elevation cross-sections \edit{perpendicular to the channel} at random locations in the region of interest. However, river shapes and hillslope topography are highly irregular\edit{, and therefore instead of taking straight cross-sections, we will investigate the topography profile along the line of the steepest descent}. For this purpose, we use the streamlines previously generated to estimate catchment width ($L_x^\text{stream}$ in \cref{sec:catchment_size}). We estimate the mean gradient for each catchment by dividing the total elevation difference for all streamlines by their total length; this is equivalent to calculating the average of a slope over all streamlines weighted by their length. Note that this method is not heavily affected by very high local gradients (\emph{e.g.} if cliffs are present in a given catchment), which would be the case if an arithmetic mean of the gradient for each streamline was calculated. We obtained values of $S_x$ ranging from as low as $0.01$ in the lowlands up to $0.3$ to $0.45$ in some highlands and mountainous regions.

\subsubsection{Estimating the gradient along the river, $S_y$}
\label{sec:estimating_Sy}

\noindent The gradient $S_y$ can be estimated similarly to $S_x$, but the points are taken along the river streams instead of streamlines. We thus estimate $S_y$ by dividing the total elevation difference across all channels in the \textit{OS Open Rivers} dataset by their total length; this method is equivalent to taking a weighted average of the slope for each individual channel, weighted by their length. The typical values of $S_y$ range from as low as $0.0005$ in the lowlands up to as high as $0.1$ in highlands and mountainous regions.

\subsection{Soil and rock properties (\texorpdfstring{$L_z$}{}, \texorpdfstring{$K_s$}{}, \texorpdfstring{$\alpha_\mathrm{MvG}$}{} \texorpdfstring{$\theta_s$}{}, \texorpdfstring{$\theta_r$}{}, \texorpdfstring{$n$}{})}

\label{sec:soil_rock_propeties}

\noindent In this section, we focus on determining the hydraulic properties of soil and rock, which are important to estimate the saturated and relative hydraulic conductivities, $K_s$ and $K_r$, appearing in~\eqref{eq:Richards}.

\begin{figure}
    \centering
    \includegraphics{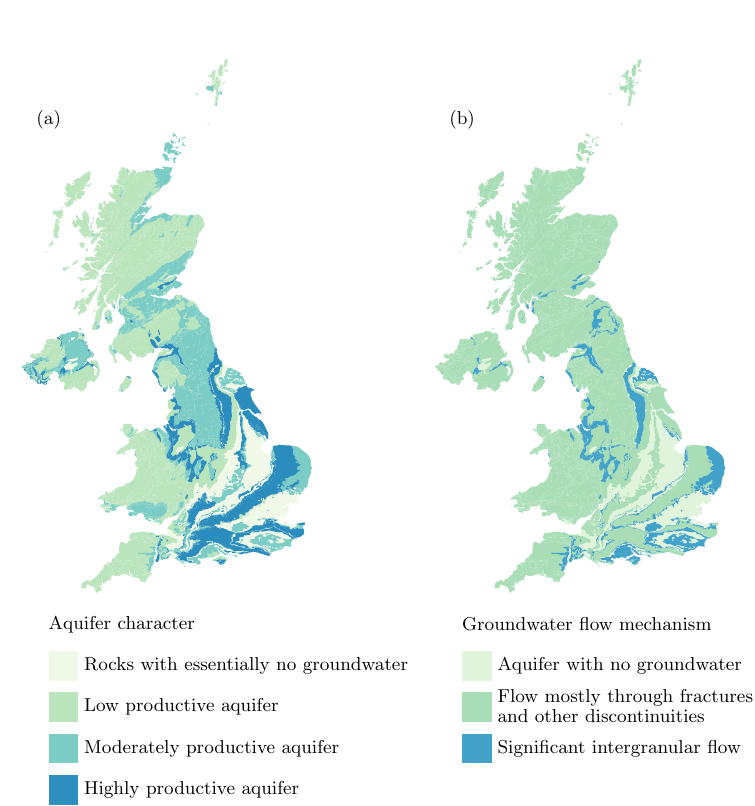}
    \caption{The hydrodynamic properties of aquifers in the UK as provided by the 625K digital hydrogeological map by the British Geological Survey (BGS). Map (a) presents the productivity of the aquifer, while map (b) presents the dominating mechanisms for the groundwater flow.}
    \label{fig:hydrogeology_map}
\end{figure}

The geological structure and hydrological properties of soils differ significantly across the UK. In some areas, such as highly productive chalk aquifers in South-East England, the soil has a very high conductivity, and almost all the rainwater reaches the groundwater table. On the other end of the spectrum, there are aquifers with essentially no groundwater, where the entire rainfall reaches the river either as subsurface flow through the soil or forms an overland flow. Based on the \emph{625K digital hydrogeological map of the UK} developed by the British Geological Society (BGS), 15\% of the UK area is classified as a highly productive aquifer, 26\% as moderately productive, 47\% as low productive, and the remaining 12\% as rock with essentially no groundwater. The geographical distribution of aquifers of different productivity levels is presented in \cref{fig:hydrogeology_map}a.

\subsubsection{Estimating the soil depth, $L_z$}

\noindent The typical depth of the conductive layer of soil can be estimated based on the UK3D dataset, which was constructed by the British Geological Survey based on measurements from 372 deep boreholes~\citep{waters2016construction}. The authors of this dataset interpolated vertical profiles, which were then interpolated to form a national network, or 'fence diagram model', of bedrock geology cross-sections. Each segment of a cross-section is assigned to one of the following qualitative aquifer designations (see \citealt{aquifer_designations}):

\begin{enumerate}[label={(\roman*)},leftmargin=*, align = left, labelsep=\parindent, topsep=3pt, itemsep=2pt,itemindent=0pt]
    \item \emph{Principal Aquifers:} layers with high intergranular and/or fracture permeability that can support river base flow on a strategic scale.
    \item \emph{Secondary Aquifers A:} permeable layers capable of forming an important source of base flow at a local rather than strategic scale.
    \item \emph{Secondary Aquifers B:} predominantly lower permeability layers.
    \item \emph{Secondary Undifferentiated:} assigned in cases where it has not been possible to attribute either category A or B to a rock type.
    \item \emph{Unproductive Strata:} layers with low permeability and negligible significance for river base flow.
\end{enumerate}

\begin{figure}
    \centering
    \includegraphics{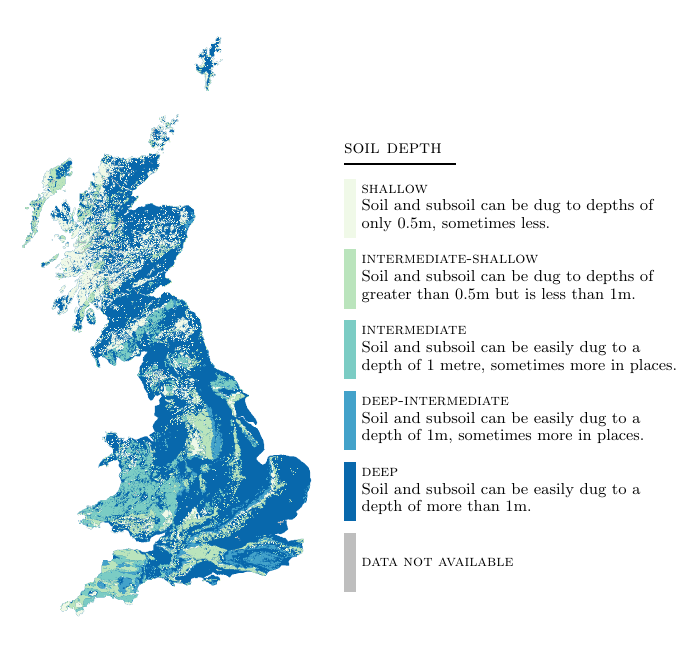}
    \caption{Thickness of soil according to the BGS Soil Parent Material Model, which provides a wide range of physical and chemical characteristics of the top layer of soil over the UK.}
    \label{fig:soil_thickness_map}
\end{figure}

\noindent We estimate the typical depth of each layer by calculating the total area of each type of rock layer and dividing it by the total length of the cross-sections, which are available in the dataset (approximately $20,000$~km). More details about the data format and our processing methods are presented in \cref{app:extracting_aquifer_thickness}. Our analysis yields the following mean thicknesses, rounded to the nearest metre, for the different types of aquifers presented in the classification (i)--(v) above: 
\begin{gather*}
\textrm{Principal} = 405~\mathrm{m}, \qquad 
\textrm{Secondary A} = 946~\mathrm{m}, \qquad 
\textrm{Secondary B} = 946~\mathrm{m}, \\
\textrm{Secondary Undiff.} = 132~\mathrm{m}, \qquad 
\textrm{Unproductive} = 75~\mathrm{m}.
\end{gather*}
Thus, the mean principal aquifer thickness is $L_z = 405$~m. However, in some catchments in the UK, this layer may be much smaller---even practically reaching zero thickness, when the groundwater flow is insignificant (notice the classification of \textit{Aquifer with no groundwater} sites in \cref{fig:hydrogeology_map}b). Therefore, such catchments need to be studied separately. In these low-productive aquifers, the subsurface flow takes place dominantly through the thin layer of soil confined from the bottom by impenetrable bedrock. Detailed quantitative data on soil thickness is not available; however, according to the maps shared by the UK Soil Observatory~\citep{lawley2009soil}, over a half of the UK consists of deep soils with thickness exceeding 1 metre, and nearly half of the soils are less than 1 metre thick (see \cref{fig:soil_thickness_map}). Therefore, in the limiting case of catchments with no groundwater, we propose  $L_z=1$~m as a reasonable choice for the typical soil thickness in regions with shallow impenetrable bedrock.

\subsubsection{Estimating the Maulem-van Genutchen (MvG) parameters}

\begin{figure}
    \centering
    \includegraphics{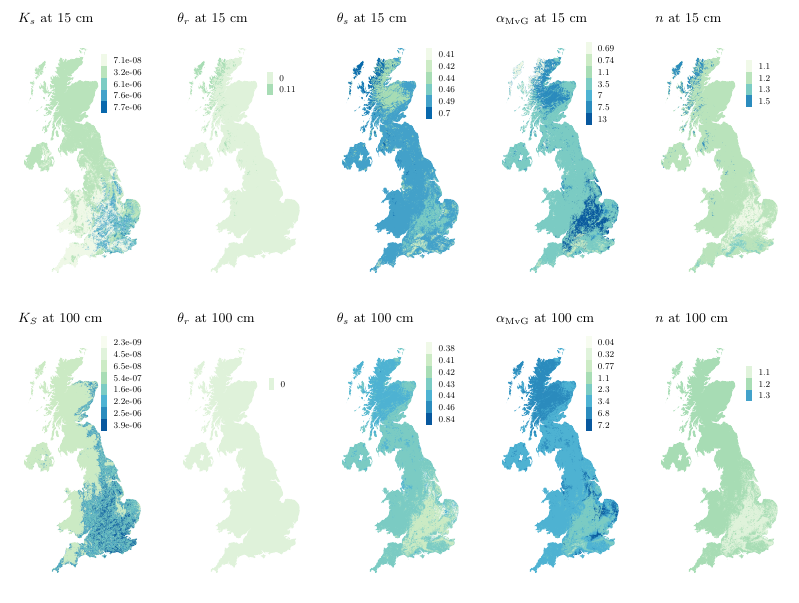}
    \caption{Mualem-Van Genuchten model parameters (from left $K_S$, $\theta_R$, $\theta_S$, $\alpha_\mathrm{MvG}$, and $n$) at a depth of 15~cm (top row) and 1~m (bottom row) according to the 3D Soil Hydraulic Database Europe. As the legend indicates, each parameter in this dataset can only have one of a few discrete values.}
    \label{fig:MvG_parameters_map}
\end{figure}

\noindent The necessary quantitative data on rock hydraulic properties (hydraulic conductivity and MvG parameters) would ideally be estimated directly through detailed experiments. However, they can also be calculated indirectly based on known soil composition. The European Soil Data Centre (ESDAC) shares the 3D Soil Hydraulic Database Europe, which uses European pedotransfer functions to estimate all MvG parameters~\citep{toth20173d}. The data is available at $1$~km and $250$~m resolutions for seven different depths (0, 5, 15, 30, 60, 100, and 200~cm). However, these estimates are not very precise since each property in the original database takes only one of a few discrete values, corresponding to specific soil types (see \cref{fig:MvG_parameters_map}). In order to provide a single estimate of these parameters, we use data corresponding to $30$ cm and extract their mean value for individual catchments. 

\subsubsection{Estimating hydraulic conductivity $K_s$}
\label{sec:estimating_Ks}

\noindent The values of the conductivity from the 3D Soil Hydraulic Database, which range from around $10^{-8}$ to $10^{-6}$~m/s (or approx. $0.001-0.1$~m/day). However, as we shall discuss below, the direct measurements conducted in several highly-productive chalk aquifers in England give higher estimates, likely because of the flow through macropores and fractures, which dominates over the intergranular flow in many areas in the UK (see \cref{fig:hydrogeology_map}b). Many studies, therefore, focus not only on analysing the hydraulic conductivity of the rock matrix, but also the bulk hydraulic conductivity, which includes the effect of both micropores forming the matrix and naturally occurring macropores/fractures. 

\begin{table}
    \centering
    \begin{tabular}{rcccc}
        {\scshape aquifer type} & {\scshape field} $K_s$ & {\scshape lab} $K_s$ & {\scshape ratio of field/lab} & {\scshape reference} \\
        Permian breccia & $1$ & $1\cdot 10^{-3}$ & $1000$ & \cite{robins1988characteristics} \\
        Cretaceous chalk & $5$ & $4\cdot 10^{-3}$ & $1250$ & \cite{price1996introducing} \\
        Triassic sandstone & $2.2$ & $1.2$ & $1.8$ & \cite{allen1998fracturing}
    \end{tabular}
    \caption{Comparison of laboratory (matrix) and field (fractures and matrix) hydraulic conductivity [m/day] for three different types of aquifers discussed by \citep{robins2006dumfries}. Field and lab values can greatly differ.}
    \label{tab:matrix_vs_fracture_K}
\end{table}

In \cref{tab:matrix_vs_fracture_K}, we cite several prior studies that have shown that the ratio of field-to-lab hydraulic conductivity values can be even greater than 1:1000. For instance, measurements conducted by \cite{robins1988characteristics} on several boreholes in the Permian and Triassic aquifers in south-west Scotland showed that hydraulic conductivity measured in the field varies from $0.1$~m/day (Solway basin) to $20$~m/day (Dumfries basin). In the study of chalk aquifers in the South Down, \cite{jones1999chalk} measured hydraulic conductivity values ranging from $0.15$ to $0.67$~m/s. \cite{gardner1990hydrology} observed that the hydraulic conductivity of English chalk aquifers increases from $1-6$~mm/day for low hydraulic potential, up to $100-1000$~mm/day, as the potential is increased above $5$kPa. This large increase in conductivity is caused by the fractures becoming saturated after increasing the potential. 

Therefore, for highly conductive aquifers, we take typical values of conductivity ranging from $K_s=10^{-6}$~m/s ($0.09$~m/day) to $K_s=10^{-4}$~m/s ($9$~m/day), but these values can be significantly lower in low-productive aquifers. In the limiting scenario of catchments with no groundwater flow, the typical conductivity is given by the conductivity of the top layer of soil.

According to \cite{kirkham2014principles}, the hydraulic conductivity of natural soils can vary from $0.05$~m/day for a clay to $30$~m/day for a silty clay loam. The variation is higher for disturbed soil materials, ranging from $0.02$~m/day for silt and clay to $600$~m/day for gravel. Therefore, for the soil, we may assume the same range. Even though the variation is higher than in the aforementioned studies for rock bulk conductivity, the typical values remain similar -- between $10^{-6}$ and $10^{-4}$~ms$^{-1}$.

\subsection{Water balance terms (\texorpdfstring{$R$}{}, \texorpdfstring{$Q$}{}, \texorpdfstring{$E$}{})}

\label{sec:water_balance}

\noindent The source term of the Saint Venant equation for the overland flow~\eqref{eq:St_Venant_overland} can be found by estimating the mean value of terms included in the water balance. It is given by:
\begin{equation}
    R=Q-E-\Delta S
\end{equation}
where $R$ is precipitation over a given catchment, $Q$ is river flow (runoff) at the outlet, $E$ is total evapotranspiration, and $\Delta S$ is the change in the water volume stored within the catchment (\emph{e.g.} in groundwater and reservoirs). Here, we not only estimate the values of $R$, $Q$, and $E$ for the UK, but also their annual peak values, which can provide a good parametrisation to model extreme precipitation events (and are used in many flood estimation models, \emph{e.g.} by \citealt{kjeldsen2008improving}). We define these quantities in terms of the mean flow per unit area of the catchment (measured in $\text{ms}^{-1}$).

The precipitation and river flow data for all UK catchments are available in the National River Flow Archive (NRFA). We use the Standardised Annual Average Rainfall (1961-90) to estimate $P$, and the mean of Gauged Daily Flow to estimate $Q$ (divided by the catchment's area, also provided by NRFA). By averaging all terms over a long period of time (several years), we should expect $\Delta S$ to tend to $0$, which allows us to estimate the mean actual evapotranspiration as $E=Q-R$.

\begin{figure}
    \centering
    \includegraphics[trim=0 0 140 0,clip]{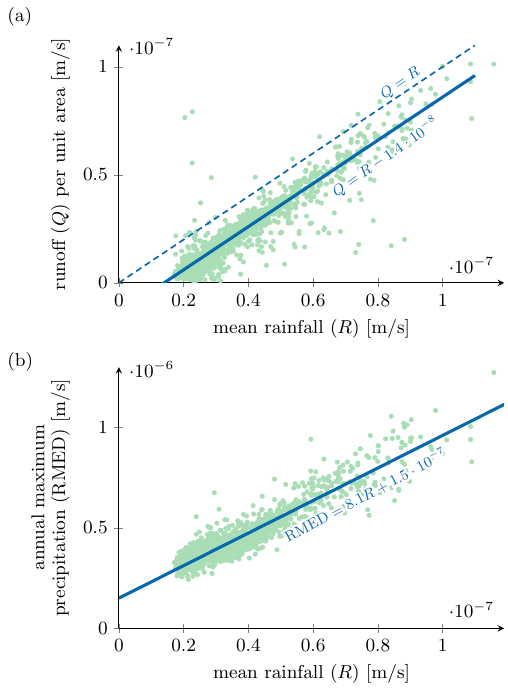}
    \caption{Dependencies between water balance terms obtained from the National River Flow Archive. The top plot (a) shows the relationship between the mean rainfall, $R$, and the mean runoff, $Q$. The difference, $R-Q$, which is a result of evapotranspiration, is approximately constant, $E \approx 1.4\cdot 10^{-8}\text{ms}^{-1}$ for the UK catchments. The bottom plot (b) show the relationship between the median of the annual maximum rainfall, RMED, and the mean rainfall, $R$. The thick solid lines in both plots correspond to the line of best fit; in (b) the fitted line confirms that RMED scales approximately linearly with $R$.}
    \label{fig:saar_vs_gdf_vs_rmed}
\end{figure}

The relationship between $Q$ and $R$ is presented in \cref{fig:saar_vs_gdf_vs_rmed}a. The mean runoff is smaller than the mean rainfall, as expected (especially for catchments with low mean rainfall). However, there are some outliers, especially among dry catchments. They correspond to situations where the mean gauged river flow exceeds the mean rainfall rate in a given catchment. Several possible reasons for this observation include the inflow of water from neighbouring catchments, a long-term trend of decreasing groundwater storage ($\Delta S$), or the mismatch between the time for which the precipitation and runoff data were available (in some catchments, the runoff data is available only for a limited period).

Based on this graph, we can deduce that the evapotranspiration is approximately independent of the other parameters $R$ and $Q$. Rainfall $R$ values vary from $10^{-8}$~m/s to $10^{-7}$~m/s, runoff $Q$ from $10^{-9}$ to $10^{-7}$, while mean evapotranspiration $E$ is usually around $1-1.5\cdot 10^{-8}$~m/s.

Finally, we estimate annual peak values of precipitation, as a measure of how intensive rainfall should be considered in rainfall-runoff models. Following \cite{faulkner1998mapping}, the standard quantity we can use is the median of annual maximum precipitation time series, RMED. As shown in \cref{fig:saar_vs_gdf_vs_rmed}b, it is closely correlated with mean precipitation $R$, and is approximately $8-10$ times higher.

\subsection{Manning's coefficients (\texorpdfstring{$n_s$}{}, \texorpdfstring{$n_c$}{})}

\label{sec:mannings_coefficient}

\noindent Manning's roughness coefficient, which appears in the overland flow equation of~\eqref{eq:Manning}, is not measured directly at the catchment scale, but is either obtained by physical model calibration or in controlled small-scale experiments. Therefore, we take typical values from engineering tables.

The Manning's coefficient of the surface, $n_s$, depends on the type of the surface~\cite{chow1959open, brunner2016hec}. The NRFA includes information about the area of each catchment covered by arable lands and grasslands ($n_s\approx 0.035~\mathrm{sm}^{-1/3}$), mountains ($n_s\approx 0.025~\mathrm{sm}^{-1/3}$), urban areas ($n_s\approx 0.015~\mathrm{sm}^{-1/3}$), and woodlands ($n_s\approx 0.16~\mathrm{sm}^{-1/3}$). Following these estimates, for each investigated catchment, we calculated a mean value of $n_s$ weighted by the area covered by each of the above terrain types.

Similarly, Manning's coefficient for the channel, $n_c$, depends on the type of the channel. Since there is no UK-wide database on channels types, we took the typical range of $n_c$ values from \cite{chow1959open}. They vary from $n_c=0.01~\mathrm{sm}^{-1/3}$ (for smooth cement channels) to $n_c=0.1~\mathrm{sm}^{-1/3}$ (for natural channels with very weedy reaches), with typical values around $n_c=0.035~\mathrm{sm}^{-1/3}$.

\subsection{Channel dimensions (\texorpdfstring{$w$}{}, \texorpdfstring{$d$}{})}

\label{sec:channel_dimensions}

\noindent Interestingly, estimating the typical channel dimensions (width $w$ and depth $d$), which appear in Manning's law for the channel flow~\eqref{eq:Manning_channel}, turns out to be quite challenging. Firstly, there is no database that provides a complete set of data on channel dimensions for UK rivers. Secondly, channel dimensions vary greatly depending on the river discharge, which can itself vary greatly depending on the catchment (though mainly on the catchment area and average rainfall). Instead, we perform the analysis via an indirect route.

Our central reference is the work by \cite{nixon1959study}, who measured the width and depth of channels of several major rivers at 29 gauging stations in England and Wales. Values ranged from a width of $w=35$~ft ($11$~m) and a depth of $d=3.1$~ft ($0.94$~m) for the River Blackwater at the Shallowfield station to a width of $w=257$~ft ($78$~m) for the Thames at the Kingston Day's Weir station, and a depth of $d=16.37$~ft ($4.99$~m) for the Wye River at the Cadora station. The data, together with similar measurements done for American rivers, were used to fit power laws describing the dependence between the channel dimensions and the full-bank discharge $Q$. \edit{From \cite{nixon1959study}, the derived power laws are: 
\begin{subequations}
\begin{align}
    \label{eq:channel_power_laws}
    d=C_1Q^{1/3}\quad\text{with}\quad & C_1=0.545\:\mathrm{s}^{1/3},\\
    w=C_2Q^{1/2}\quad\text{with}\quad & C_2=1.65\:\mathrm{s}^{1/2}\mathrm{ft}^{-1/2}=3.00\:\mathrm{s}^{1/2}\mathrm{m}^{-1/2},
\end{align}
\end{subequations}
where $Q$ is the full-bank discharge, while $C_1$ and $C_2$ are fitted constants. These scaling laws can also be justified by analysis of mechanical equilibrium of the sediment bed, see \emph{e.g.} \cite{henderson1963stability} and \cite{devauchelle2011longitudinal}.} 
    
\edit{Because data on channel dimensions is not available for most UK catchments, we use the above power laws to estimate the channel dimensions at the outlet of each studied UK catchment. We note that this is a reasonable approach: \cite{nixon1959study} derived such laws based on empirical relations for a representative sample of UK catchments. Equally, \cite{wharton1992flood} used similar scaling laws to estimate the peak annual flows in 70 UK catchments based on the channel dimensions.}

\subsection{A summary of results on characteristic values}
\label{sec:results}

\noindent Table \ref{tab:parameter_summary} summarises the typical values of catchment model parameters for all gauged catchments in the UK, with boundaries defined as in the National River Flow Archive. A sample of raw data and their spatial distributions is included in \cref{app:spatial_distribution}. In cases where no spatial data was available, an estimate from the literature was obtained (these entries are marked with $\dagger$). Where available estimates obtained with more than one method are compared. \edit{The median value of the parameters presented in \cref{tab:parameter_summary} will be used in Part 2 to formulate a simple benchmark scenario characterised by similar physical properties to the UK catchments.}

\afterpage{
\clearpage
\begin{landscape}
    \begin{table}
        \tiny
        \centering
        \begin{tabular}{rcccccccc}
            \textsc{parameter} & \textsc{symbol} & \textsc{unit} & \textsc{25\% quantile} & \textsc{median value} & \textsc{75\% quantile} & \begin{tabular}{@{}c@{}}\textsc{data source /} \\ \textsc{datasets used} \end{tabular} & \begin{tabular}{@{}c@{}}\textsc{Example MIKE SHE calibration} \\ \citep{doummar2012simulation} \end{tabular} & \begin{tabular}{@{}c@{}}\textsc{Benchmark scenarios} \\ \citep{maxwell2014surface}\end{tabular} \\
            
            \multirow{2}{*}{hillslope width} & $L_x^\mathrm{area}$ & \multirow{2}{*}{m} & $492$ & $683$ & $1209$ & \begin{tabular}{@{}c@{}}NRFA \\ OS VectorMap\end{tabular}  & \multirow{2}{*}{approx. 1200} & \multirow{2}{*}{\begin{tabular}{@{}c@{}}400 (hillslope)\\800 (V-shaped)\end{tabular}} \\
            & $L_x^\mathrm{stream}$ & & $528$ & $616$ & $710$ & \begin{tabular}{@{}c@{}}OS Terrain 50 \\ OS VectorMap\end{tabular} & & \vspace{2 mm} \\
            
            \multirow{4}{*}{catchment length} & $L_y^\text{long}$ & \multirow{4}{*}{m} & $13000$ & $22800$ & $38300$ & OS Open Rivers & \multirow{4}{*}{approx. 39000} & \multirow{4}{*}{1000 (V-shaped)} \\
            & $L_y^\text{all}$ & & $36500$ & $130000$ & $366000$ & OS VectorMap & &
            \vspace{2 mm} \\
            & $L_y^\text{main}$ & & $25200$ & $71600$ & $204000$ & OS Open Rivers & &
            \vspace{2 mm} \\
            & $L_y^\text{trib}$ & & $725$ & $945$ & $1212$ & OS Open Rivers & &
            \vspace{2 mm} \\
            
            aquifer thickness & $L_z$ & m & $159$ & $684$ & $1550$ & UK3D  & unknown & -- \vspace{2 mm} \\
            soil thickness & $L_z^*$ & m & & $\sim 1$\textsuperscript{$\dagger$} & & BGS Soil property data & unknown & 5 (hillslope) \vspace{2 mm} \\
            slope along the river & $S_y$ & $-$ & $0.00504$ & $0.014$ & $0.0316$ & \begin{tabular}{@{}c@{}}OS Terrain 50 \\ OS Open Rivers\end{tabular}  & unknown & \begin{tabular}{@{}c@{}}0.02 (V-shaped)\\0 (hillslope)\end{tabular} \\
            slope along the hillslope & $S_x$ & $-$ & $0.0386$ & $0.075$ & $0.131$ & \begin{tabular}{@{}c@{}}OS Terrain 50 \\ OS VectorMap\end{tabular}  & unknown & \begin{tabular}{@{}c@{}}0.05 (V-shaped)\\0.0005 (hillslope)\end{tabular} \vspace{2 mm} \\
            \multirow{2}{*}{saturated soil conductivity} & \multirow{2}{*}{$K_s$} & \multirow{2}{*}{ms$^{-1}$} &
             & $10^{-6}$ -- $10^{-4}$\textsuperscript{(a)}\textsuperscript{$\dagger$} &  & multiple sources & \multirow{2}{*}{$10^{-5}$ -- $10^{-2}$} & \multirow{2}{*}{\begin{tabular}{@{}c@{}}from $1.16\cdot 10^{-7}$ \\ to $1.16\cdot 10^{-5}$\end{tabular}} \\
            & & &
            $2.98\cdot 10^{-6}$\textsuperscript{(b)} & $3.2\cdot 10^{-6}$\textsuperscript{(b)} & $6.26\cdot 10^{-6}$\textsuperscript{(b)} & ESDAC & & \vspace{2 mm} \\
            MvG model $\alpha$ parameter & $\alpha_\mathrm{MvG}$ & m$^{-1}$ & $3.47$ & $3.7$ & $6.14$ & ESDAC & 0.00122 -- 0.1 & 100 \\
            residual water content & $\theta_r$ & $-$ & $0$ & $0$ & $0$ & ESDAC & 0.1-0.2 & 0.2 \\
            saturated water content & $\theta_s$ & $-$ & $0.473$ & $0.488$ & $0.491$ & ESDAC & 0.2-0.4 & 1 \\
            MvG $n$ coefficient & $n$ & $-$ & $1.19$ & $1.19$ & $1.19$ & ESDAC & 1 -- 3 & 2 \vspace{2 mm} \\
            
            precipitation rate & $R$ & ms$^{-1}$ & $2.27\cdot 10^{-8}$ & $2.95\cdot 10^{-8}$ & $4.2\cdot 10^{-8}$ & NRFA & $2.41\cdot 10^{-8}$ -- $3.91\cdot 10^{-8}$ & \begin{tabular}{@{}c@{}}$3\cdot 10^{-6}$ (hillslope) \\ $5.5\cdot 10^{-6}$ (V-shaped)\end{tabular} \\
            runoff per unit area & $Q$ & ms$^{-1}$ & $8.18\cdot 10^{-9}$ & $1.59\cdot 10^{-8}$ & $2.87\cdot 10^{-8}$ & NRFA & $6.44\cdot 10^{-9}$ -- $1.47\cdot 10^{-8}$ & same as rainfall \\
            evapotranspiration rate & $E$ & ms$^{-1}$ & $1.11\cdot 10^{-8}$ & $1.34\cdot 10^{-8}$ & $1.54\cdot 10^{-8}$ & NRFA & $1.80\cdot 10^{-8}$ -- $2.43\cdot 10^{-8}$ & $0$ \\
            peak annual precipitation & RMED & ms$^{-1}$ & $3.50\cdot 10^{-7}$ & $3.99\cdot 10^{-7}$ & $4.88\cdot 10^{-7}$ & NRFA & -- & -- \\
            peak annual river flow & QMED & ms$^{-1}$ & $1.14\cdot 10^{-7}$ & $2.76\cdot 10^{-7}$ & $5.10\cdot 10^{-7}$ & NRFA & -- & -- \vspace{2 mm} \\
            
            \begin{tabular}{@{}c@{}}Manning's coefficient \\ for the land surface\end{tabular} & $n_s$ & sm$^{-1/3}$ & $0.0442$ & $0.0509$ & $0.0633$ & \begin{tabular}{@{}c@{}}NRFA \\ \cite{brunner2016hec}\end{tabular} & unknown & \begin{tabular}{@{}c@{}}$0.1986$ (hillslope) \\ $0.015\cdot 10^{-6}$ (V-shaped)\end{tabular} \\
            \begin{tabular}{@{}c@{}}Manning's coefficient \\ for the channel\end{tabular} & $n_c$ & sm$^{-1/3}$ & & 0.01 -- 0.1 \textsuperscript{$\dagger$} & & \cite{chow1959open} & unknown & $0.15\cdot 10^{-6}$ (V-shaped) \vspace{2 mm} \\
            channel's width & $w$ & m & & 11 -- 78 \textsuperscript{$\dagger$} & & \cite{nixon1959study} & unknown & \begin{tabular}{@{}c@{}}0 (hillslope) \\ 20 (V-shaped)\end{tabular} \\
            channel's depth at outlet & $d$ & m & & 0.94 -- 4.99 \textsuperscript{$\dagger$} & & \cite{nixon1959study} & unknown & 4 (hillslope) \\
        \end{tabular}
        \vspace{1mm}
        \begin{tabular}{ll}
            (a) Hydraulic conductivity of both bedrock and soil estimated based on various studies, &
            (b) Hydraulic conductivity of the soil as given in Hydraulic Database of Europe by ESDAC
        \end{tabular}
        \vspace{-0.4mm}
        \begin{tabular}{l}
        $\dagger$ Numbers represent a range of values obtained from the literature; These quantities are not estimated individually on the catchment level.
        \end{tabular}
        \caption{Summary of catchment model parameters values for UK catchments.}
        \label{tab:parameter_summary}
    \end{table}
\end{landscape}
}

Based on \cref{tab:parameter_summary}, we can make the following notes:
\begin{enumerate}[label={(\roman*)}, leftmargin=*, align = left, labelsep=\parindent, topsep=3pt, itemsep=2pt,itemindent=0pt]
\item The total stream length in the catchment is approximately twice as high when including all surface water streams and bodies from the OS VectorMap ($L_y^\text{all}$) than when looking only at the main rivers, which are included in OS Open Rivers dataset ($L_y^\text{main}$), cf. \cref{fig:river_datasets_comparison}. The longest stream ($L_y^\text{long}$) typically contributes to from 20\% to 50\% of the total length of the rivers in OS Open River, with a higher percentage corresponding to shorter rivers (with fewer and shorter tributaries). The estimated lengths greatly vary as a result of the wide range of gauged catchments areas used in the study. The exception is the distance between tributaries ($L_t^\text{trib}$) ranging from $725$ to $1212$~m. As an intensive quantity, this quantity does not scale with the catchment area. 

\item Hillslope width estimates in the majority of catchments are consistent between the two proposed methods. A difference can be observed in catchments with the lowest density of streams, for which streamlines turn out to be much shorter than the width between streams. \edit{The reason is that the data includes numerous streamlines that start far from the river, but terminate at a local elevation minimum, and hence do not reconnect with the river. In our study, such streamlines cover over $18\%$ of the UK's area. We do not account for such streamlines when estimating the hillslope width.}

\item Table \ref{tab:parameter_summary} provides an indication that the slope along the hillslope $S_x$ is typically much larger than the slope along the channel $S_y$, which is a general feature of the topography of river valleys. \edit{By the two-dimensional Manning's equation \eqref{eq:Manning}, the flow follows the direction of the steepest descent. Therefore, we would then expect that the rainfall water is transferred towards the main river through streams mainly directed along the hillslopes (with $S_x$ slope). Only after it reaches the channel, it is transferred down the river with $S_y$ slope.} The consequences of this observation will be explored in detail in Part 2.

\item The highest disagreement can be observed between the soil conductivity as reported in the ESDAC dataset and the literature. The latter measured the bulk soil conductivity of the soil (and bedrock), which is a better macroscopic estimate than the one provided in the ESDAC dataset.

\item The last two columns of \cref{tab:parameter_summary} include example values from the literature. The first column includes parameter ranges suggested by~\cite{doummar2012simulation} to model a real-world Karst catchment in Southwest Germany. The catchment width is higher than typical values for the UK. Notable differences between the suggested parameter ranges and estimations for UK catchments are in soil conductivity, where higher values come from measurements of observed flow velocities using artificial tracers in highly conductive areas of the investigated system. These velocities can be very high in well-fractured Karst systems, as is also observed in some measurements in the UK [cf. the chalk aquifer of Yorkshire studied by~\cite{ward1997use}].

\item The last column includes parameter values suggested in two simple benchmark scenarios by~\cite{maxwell2014surface} for integrated catchment model intercomparison. The first scenario includes a single hillslope tilted in the $x$-direction, ending in a flat river of constant depth. Only subsurface flow in a thin 5-metre-deep soil layer is modelled. The second scenario includes a two-dimensional tilted V-shaped catchment with a different elevation gradient along and perpendicular to the river. However, this latter one is only used to model surface flow---subsurface flow is not involved.

One should note a few differences between parameter values used in these scenarios. The rainfall values are higher, since they correspond to typical intensive rainfall, not to an annual mean estimated for UK catchments. The next difference is a much lower aquifer thickness, since in benchmark scenarios, it represents only the soil layer ($L_z^*=5$ m), ignoring the subsurface water percolation to the usually much deeper permeable bedrock layer (on average $L_z=684$ m based on the UK3D dataset). Another important difference is the low soil conductivity in the scenarios: \edit{the saturated hydraulic conductivity, $K_s$, between $1.16\cdot 10^{-7}$ to $1.16\cdot 10^{-5}$ m/s is lower than the reported typical values ranging from $10^{-4}$ to $10^{-6}$ m/s (estimated based on large-scale studies in the UK, see \cref{sec:estimating_Ks}).} Note that the combination of the two previous factors (thin aquifer with low conductivity) typically leads to model results, in simple benchmarks, where the majority of land is covered with surface water even for small rainfalls (cf. modelling and numerical simulations in Part 2).
\end{enumerate}

\section{Further investigation of distribution and correlation}
\label{sec:statistical_analysis}

\noindent In the previous section, we presented the results of an extensive analysis of the typical scaling of those key parameters expected to play a role in the dynamics of hydrology at the level of a catchment site. Naturally, an important question relates to the issue of which parameters are expected to be the most important in the description of physical observations or in numerical simulations. Such questions will form the basis of our investigation in Parts 2 and 3, where we develop scaling laws involving non-dimensional parameters. In this section, we focus on a statistical interpretation of such issues, and study the distribution and correlation of parameters, as well as the subsequent catchment classification. Our analysis is divided into the application of three statistical methods: analysis of cross-correlation of parameters (\cref{sec:corr}), grouping of catchments using k-means clustering (\cref{sec:cluster}), and Principal Component Analysis for catchment properties (\cref{sec:PCA}).

\subsection{Regional variability and correlation of parameters} \label{sec:corr}

\begin{figure}
    \centering
    \includegraphics{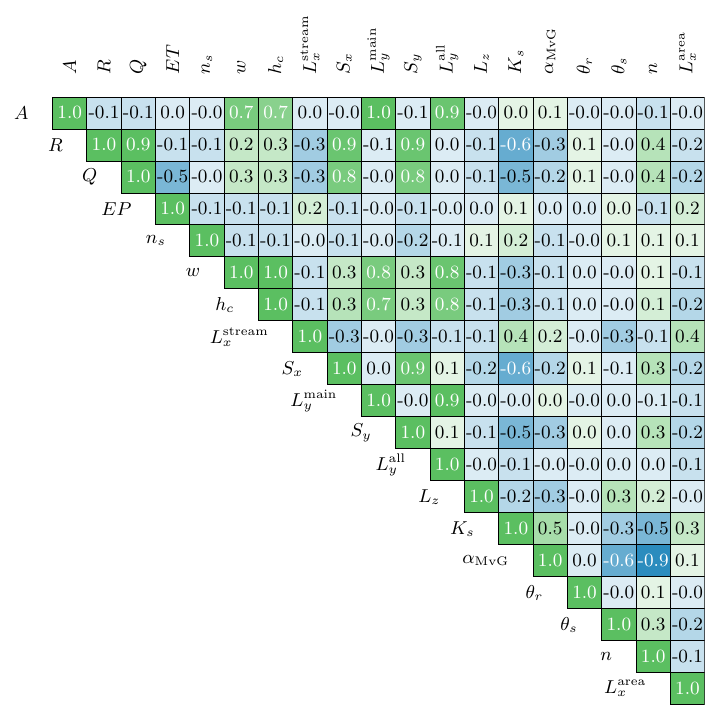}
    \caption{Correlogram of physical parameters  summarised in \cref{tab:parameter_summary}. It presents the Pearson correlation coefficient calculated based on values of the given parameters, which were estimated for UK catchments specified in the NRFA. Incomplete records were omitted.}
    \label{fig:correlation_plot}
\end{figure}

\noindent Note that the estimates presented in \cref{tab:parameter_summary} refer to mean values across the UK. However, regional averages can significantly differ\edit{, as is represented by the interquartile range}. As an example, consider the distribution of certain parameter mean values for UK catchments, as listed in the National River Flow Archives and presented in \cref{fig:UK_map_grid}. One can observe that highland areas, characterised by a higher elevation gradient (both along rivers and hillslopes), are also accompanied by higher rainfalls and lower soil conductivity (and usually low aquifer thickness). These observations are confirmed by the correlation matrix presented in the \cref{fig:correlation_plot}. The higher precipitation in mountainous regions was already observed by~\cite{duckstein1973elevation}. Its impact on the aquifer properties (such as soil capacity and infiltration rate) was used by~\cite{bell1998grid} to formulate the Grid Model and its successor, the Grid-to-Grid model~\citep{bell2007development}. The high correlation of catchment area and river length comes from the fractal structure of drainage networks~\citep{rodriguez2001fractal}, which develop to uniformly cover available space. Finally, there is a strong correlation between the $\alpha_\mathrm{MvG}$, $n$ and $\theta_S$ parameters from the Mualem-Van Genuchten model parameters, since all were estimated by~\cite{toth20173d} based on the same soil composition dataset.

\subsection{Classification of catchments using cluster analysis} \label{sec:cluster}

\noindent In order to better understand the similarities between different types of catchments, characterised by different combinations of physical parameters, we perform a cluster analysis. The goal is to construct a specified number of typical parameter combinations that reflect the spatial variation of UK catchments. Here, we use k-means clustering on the parameters summarised in \cref{tab:parameter_summary} to find a set of three clusters [see \emph{e.g.}~\cite{kaufman2009finding}]. Parameters not directly associated or extracted with specific individual catchments in mind are removed from consideration ($w$, $d$, $n_c$, $L_z^*$). Also ignored are quantities that scale with the catchment areas, since these parameters are dependent on the location of the catchment outlet, rather than the physical properties characterising a given region ($L_y^\mathrm{all}$, $L_y^\mathrm{main}$ and $L_y^\mathrm{long}$). Finally, the residual water content, $\theta_r$, was removed from consideration since it is almost always zero. This leaves us with the remaining 15 variables. Only catchments for which all these parameters are available are used for clustering.

\begin{figure}
    \centering
    \includegraphics{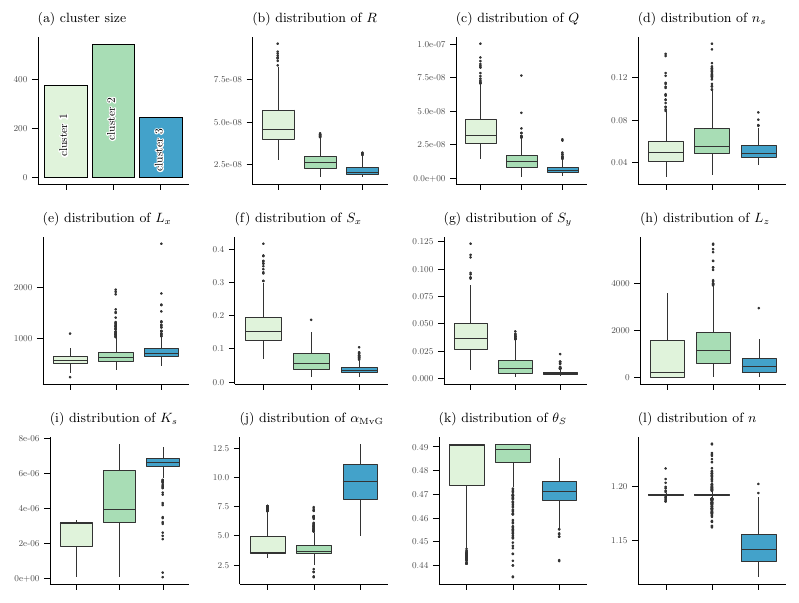}
    \caption{Number of catchments belonging to each cluster (top left) and box plots showing the distribution of parameters among the catchments belonging to each graph.}
    \label{fig:clustering_results}
\end{figure}

The distribution of parameter values for each cluster is presented in \cref{fig:clustering_results}. The figure highlights some key features of each cluster: 
\begin{enumerate}[label={(\roman*)}, leftmargin=*, align = left, labelsep=\parindent, topsep=3pt, itemsep=2pt,itemindent=0pt]
    \item From \cref{fig:clustering_results}(f) and (g), we note that the first cluster is characterised by the highest elevation gradients, which mostly represent catchments located in highlands, while the third cluster is associated with the lowest elevation gradients, representing relatively flat lowlands. Intermediate values are represented by the largest second cluster.
    \item Fig. \ref{fig:clustering_results}(i)--(l) show that the third cluster, representing the lowlands, has significantly different soil properties from the other two clusters. It has the highest hydraulic conductivity, $K_s$, and the Mualem-Van Genuchten $\alpha_\mathrm{MvG}$ parameter, as well as the lowest saturated water content, $\theta_S$, and MvG coefficient, $n$.
    \item Finally, in Fig.~\ref{fig:clustering_results}(b) and (c), we note that the first cluster, representing the highlands, has significantly higher mean precipitation, $R$, and runoff, $Q$, values, which is related to the fact that precipitation is higher in eastern parts of the UK, coinciding with many of the highland and mountainous areas.
\end{enumerate}

\subsection{Catchment characterisation using PCA}
\label{sec:PCA}

\noindent In order to better understand the separation between the class properties, one can use principal component analysis (PCA), which allows us to find the linear combination of the original parameters with the highest variation in the dataset. Table \ref{tab:PCA_table} summarises the linear coefficients (eigenvectors) representing the first four principal components. Based on the parameters with the highest eigenvalues, \edit{our intuitive interpretation of the components is as follows}. One can interpret the first principal component as strongly related to the topography of the terrain. The highland terrain with higher slopes and higher rainfall/runoff values is characterised by high values of the first principal components, while the lowlands correspond to low values. The second principal component is high for catchments with high aquifer thickness, high soil saturation, low $\alpha_\mathrm{MvG}$ and/or high Manning's $n$ roughness coefficient. Typically, such a configuration of parameters represents catchments, in which groundwater flow has a relatively high importance. \edit{On the contrary, catchments with a low value of the second component will be characterised by a higher contribution of the overland flow component.}

\begin{figure}
    \centering
    \includegraphics{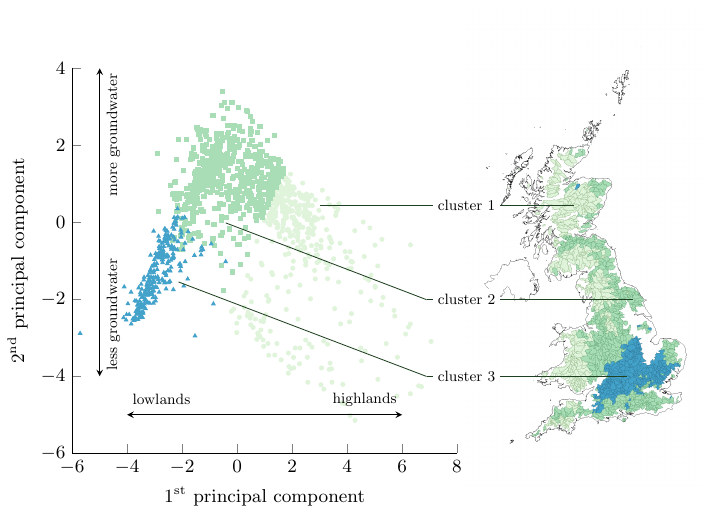}
    \caption{The illustration of the first two principal components. The graph on the left shows the values of principal components for catchments belonging to each of the three clusters. The map on the right shows the geographic distribution of these clusters.}
    \label{fig:PCA_summary}
\end{figure}

Fig. \ref{fig:PCA_summary} presents the clusters in a two-dimensional space spanned by the first two principal components. Here the separation between classes is apparent---as observed before, they vary in terrain topography. Additionally, the \edit{second} cluster represents the most productive aquifers with the highest importance of groundwater. In the other two classes, the lower importance of groundwater is balanced by higher overland flows. \editt{In the case of cluster 1, the groundwater flow is limited by low hydraulic conductivity of the soil (cf. \cref{fig:clustering_results}i), while in cluster 3, it is limited by the very flat terrain (cf. \cref{fig:clustering_results}f-g), which can easily become saturated.} 

\begin{figure}
    \centering
    \includegraphics[trim=0 0 0 30,clip]{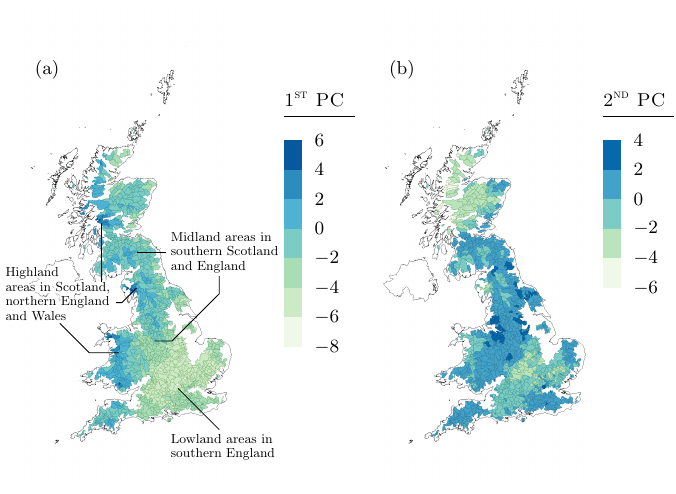}
    \caption{Spatial distribution of values of the first (a) and second (b) principal components of the catchment parameters. Note that visually the value of the first component seems to coincide with the main lowland, midland, and highland regions in the UK.}
    \label{fig:PCA_components}
\end{figure}

Fig. \ref{fig:PCA_components} also illustrates the spatial distribution of the first two principal components, as well as the geographical distribution of catchments belonging to different classes. \edit{Based on visual inspection of the distribution of regions associated with different values of the first principal component in \cref{fig:PCA_components}a, the measure seems to be highly spatially autocorrelated. This is due to the UK's distinct lowland, midland and highland regions, characterised by low, medium, and high values of the first principal component, respectively.} The spatial autocorrelation of the second principal component can also be observed in \cref{fig:PCA_components}b; however, this component varies much more locally---we expect it to be dependent on the local geological structure of the aquifer and the local soil composition.

\section{Discussion}
\label{sec:discussion}

\noindent In order to produce minimal mathematical models of surface-subsurface flows in catchments, a crucial step is to establish the typical size and distribution of the key parameters, which has been the main goal of this work. As we have noted, a comprehensive treatment of this topic does not seem to have been previously undertaken in the style we have presented here.

In this paper, we compiled a set of data processing methods that allow the extraction of typical values of physical parameters used in PDE-based catchment models. The methods were applied to extract properties of UK catchments based on publicly available datasets from the National River Flow Archive, European Soil Data Centre, British Geological Society and Ordnance Survey. The parameter values for each catchment and the source code used to obtain them are published in our GitHub repository~\citep{github1} for use by other researchers and practitioners. The key values of the parameters are summarised in \cref{tab:parameter_summary}.

Our application of cluster analysis on the collected data sets allowed us to further classify those groupings of parameters that determine different classes of catchments. The intuitive interpretation of our findings seems reasonable: a key clustering component relates to those parameters characterising the terrain topography, while the second component is related to the groundwater importance in the runoff generation. For the demonstration purposes, UK catchments were divided into three clusters, each characterised by distinctive values of these principal components. Based on \cref{fig:PCA_summary}, we concluded that the clusters approximately represent lowland, midland, and highland areas, with midland areas often characterised by the highest groundwater importance in water transfer.

The presented results can be used for verification of parameter values calibrated by numerical catchment models and for preparing realistic benchmark scenarios. Ideal catchment model(s) should thus provide quantitative or qualitative agreement with catchments belonging to any of the identified clusters. We expect that this is also an important criterion when creating and validating simplified (conceptual and statistical) catchment models, which often have a tendency to become overfitted to the training data~\citep{beven2018hypothesis,beven2019make}. Identifying scenarios where a given model may fail can lead to further model improvement. 

We shall continue our work in Parts 2 and 3, where the parameter estimates are crucial in allowing for the development and analysis of physical models of surface and subsurface flows. 

\mbox{}\par
\noindent \textbf{Acknowledgements.} We thank Sean Longfield (Environmental Agency) for many useful interactions and for motivating this work via the 7th Integrative Think Tank hosted by the Statistical and Applied Mathematics CDT at Bath (SAMBa). We also thank Thomas Kjeldsen (Bath), Tristan Pryer (Bath), and Rob Lamb (Lancaster/JBA Trust) for insightful discussions. We are indebted to the reviewers and the JFM editorial team---their comments and suggestions were instrumental in the final development of this paper. Piotr Morawiecki is supported by a scholarship from the EPSRC Centre for Doctoral Training in Statistical Applied Mathematics at Bath (SAMBa), under the project EP/S022945/1.

\mbox{}\par
\noindent \textbf{Declaration of Interests.} The authors report no conflict of interest.

\bibliographystyle{plainnat}
\bibliography{bibliography}

\begin{thebibliography}{68}
\providecommand{\natexlab}[1]{#1}
\providecommand{\url}[1]{\texttt{#1}}
\expandafter\ifx\csname urlstyle\endcsname\relax
  \providecommand{\doi}[1]{doi: #1}\else
  \providecommand{\doi}{doi: \begingroup \urlstyle{rm}\Url}\fi

\bibitem[Abbott et~al.(1986)Abbott, Bathurst, Cunge, O'Connell, and
  Rasmussen]{abbott1986introduction1}
M.~B. Abbott, J.~C. Bathurst, J.~A. Cunge, P.~E. O'Connell, and J.~Rasmussen.
\newblock An introduction to the {European Hydrological System—Systeme
  Hydrologique Europeen}, "{SHE}", 1: History and philosophy of a
  physically-based, distributed modelling system.
\newblock \emph{J. Hydrol.}, 87\penalty0 (1-2):\penalty0 45--59, 1986.

\bibitem[Allen et~al.(1998)Allen, Bloomfield, Gibbs, and
  Wagstaff]{allen1998fracturing}
D.~J. Allen, J.~P. Bloomfield, B.~R. Gibbs, and S.~J. Wagstaff.
\newblock Fracturing and the hydrogeology of the {Permo-Triassic} sandstones in
  {E}ngland and {W}ales.
\newblock Technical report, British Geological Survey, 1998.

\bibitem[Anderson et~al.(2015)Anderson, Woessner, and
  Hunt]{anderson2015applied}
M.~P. Anderson, W.~W. Woessner, and R.~J. Hunt.
\newblock \emph{Applied groundwater modeling: simulation of flow and advective
  transport}.
\newblock Academic press, 2015.

\bibitem[Bear(1972)]{bear1972dynamics}
J.~Bear.
\newblock \emph{Dynamics of Fluids in Porous Media}.
\newblock Dover Publications, 1972.

\bibitem[Bell and Moore(1998)]{bell1998grid}
V.~A. Bell and R.~J. Moore.
\newblock A grid-based distributed flood forecasting model for use with weather
  radar data: {P}art 1. {F}ormulation.
\newblock \emph{Hydrol. Earth Syst. Sci.}, 2\penalty0 (2/3):\penalty0 265--281,
  1998.

\bibitem[Bell et~al.(2007)Bell, Kay, Jones, and Moore]{bell2007development}
V.~A. Bell, A.~L. Kay, R.~G. Jones, and R.~J. Moore.
\newblock Development of a high resolution grid-based river flow model for use
  with regional climate model output.
\newblock \emph{Hydrol. Earth Syst. Sci.}, 11\penalty0 (1):\penalty0 532--549,
  2007.

\bibitem[Beven(2018)]{beven2018hypothesis}
K.~Beven.
\newblock On hypothesis testing in hydrology: {W}hy falsification of models is
  still a really good idea.
\newblock \emph{Wiley Interdiscip. Rev.: Water}, 5\penalty0 (3):\penalty0
  e1278, 2018.

\bibitem[Beven(2019)]{beven2019make}
K.~Beven.
\newblock How to make advances in hydrological modelling.
\newblock \emph{Hydrol. Res.}, 50\penalty0 (6):\penalty0 1481--1494, 2019.

\bibitem[Beven(2011)]{beven2011rainfall}
K.~J. Beven.
\newblock \emph{Rainfall-runoff modelling: the primer}.
\newblock John Wiley \& Sons, 2011.

\bibitem[Beven and Kirkby(1977)]{beven1977towards}
K.~J. Beven and M.~J. Kirkby.
\newblock \emph{Towards a simple, physically-based, variable contributing area
  model of catchment hydrology}.
\newblock University of Leeds, School of Geography, 1977.

\bibitem[Bl{\"o}schl(2006)]{bloschl2006rainfall}
G.~Bl{\"o}schl.
\newblock Rainfall-runoff modeling of ungauged catchments.
\newblock \emph{Encyclopedia of hydrological sciences}, 2006.

\bibitem[{British Geological Society}(2022)]{GB_viewer}
{British Geological Society}.
\newblock Geology of {Britain} viewer.
\newblock \url{https://www.bgs.ac.uk/map-viewers/geology-of-britain-viewer/},
  2022.
\newblock Accessed: 2022-10-26.

\bibitem[Brunner(2016)]{brunner2016hec}
G.~W. Brunner.
\newblock {HEC-RAS} river analysis system 2{D} modeling user’s manual.
\newblock \emph{US Army Corps of Engineers—Hydrologic Engineering Center},
  pages 1--171, 2016.

\bibitem[Brunner and Simmons(2012)]{brunner2012hydrogeosphere}
P.~Brunner and C.~T. Simmons.
\newblock Hydrogeosphere: a fully integrated, physically based hydrological
  model.
\newblock \emph{Groundwater}, 50\penalty0 (2):\penalty0 170--176, 2012.

\bibitem[Calver et~al.(2009)Calver, Stewart, and
  Goodsell]{calver2009comparative}
A.~Calver, E.~Stewart, and G.~Goodsell.
\newblock Comparative analysis of statistical and catchment modelling
  approaches to river flood frequency estimation.
\newblock \emph{J. Flood Risk Manage.}, 2\penalty0 (1):\penalty0 24--31, 2009.

\bibitem[Chaudhry(2007)]{chaudhry2007open}
M.~H. Chaudhry.
\newblock \emph{Open-channel flow}.
\newblock Springer Science \& Business Media, 2007.

\bibitem[Chow(1959)]{chow1959open}
V.~T. Chow.
\newblock Open-channel hydraulics, 1959.

\bibitem[Chow and Ben-Zvi(1973)]{chow1973hydrodynamic}
V.~T. Chow and A.~Ben-Zvi.
\newblock Hydrodynamic modeling of two-dimensional watershed flow.
\newblock \emph{Journal of the Hydraulics Division}, 99\penalty0 (11):\penalty0
  2023--2040, 1973.

\bibitem[Clauser(1992)]{clauser1992permeability}
C.~Clauser.
\newblock Permeability of crystalline rocks.
\newblock \emph{Eos, Transactions American Geophysical Union}, 73\penalty0
  (21):\penalty0 233--238, 1992.

\bibitem[Cole et~al.(1991)Cole, Slade, Jones, and Gregory]{cole1991reliable}
J.~A. Cole, S.~Slade, P.~D. Jones, and J.~M. Gregory.
\newblock Reliable yield of reservoirs and possible effects of climatic change.
\newblock \emph{Hydrol. Sci. J.}, 36\penalty0 (6):\penalty0 579--598, 1991.

\bibitem[Devauchelle et~al.(2011)Devauchelle, Petroff, Lobkovsky, and
  Rothman]{devauchelle2011longitudinal}
O.~Devauchelle, A.~P. Petroff, A.~E. Lobkovsky, and D.~H. Rothman.
\newblock Longitudinal profile of channels cut by springs.
\newblock \emph{J. Fluid Mech.}, 667:\penalty0 38--47, 2011.

\bibitem[Dogan and Motz(2005)]{dogan2005saturated}
A.~Dogan and L.~H. Motz.
\newblock Saturated-unsaturated 3{D} groundwater model. i: {D}evelopment.
\newblock \emph{J. Hydrol. Eng.}, 10\penalty0 (6):\penalty0 492--504, 2005.

\bibitem[Douglas and Peucker(1973)]{douglas1973algorithms}
D.~H. Douglas and T.~K. Peucker.
\newblock Algorithms for the reduction of the number of points required to
  represent a digitized line or its caricature.
\newblock \emph{Cartographica}, 10\penalty0 (2):\penalty0 112--122, 1973.

\bibitem[Doummar et~al.(2012)Doummar, Sauter, and Geyer]{doummar2012simulation}
J.~Doummar, M.~Sauter, and T.~Geyer.
\newblock Simulation of flow processes in a large scale karst system with an
  integrated catchment model ({Mike She})--identification of relevant
  parameters influencing spring discharge.
\newblock \emph{J. Hydrol.}, 426:\penalty0 112--123, 2012.

\bibitem[Duckstein et~al.(1973)Duckstein, Fogel, and
  Thames]{duckstein1973elevation}
L.~Duckstein, M.~M. Fogel, and J.~L. Thames.
\newblock Elevation effects on rainfall: {A} stochastic model.
\newblock \emph{J. Hydrol.}, 18\penalty0 (1):\penalty0 21--35, 1973.

\bibitem[EA(2017)]{aquifer_designations}
EA.
\newblock Guidance: Protect groundwater and prevent groundwater pollution.
\newblock
  \url{https://www.gov.uk/government/publications/protect-groundwater-and-prevent-groundwater-pollution/protect-groundwater-and-prevent-groundwater-pollution},
  2017.
\newblock Accessed: 2023-05-04.

\bibitem[Faulkner and Prudhomme(1998)]{faulkner1998mapping}
D.~S. Faulkner and C.~Prudhomme.
\newblock Mapping an index of extreme rainfall across the {UK}.
\newblock \emph{Hydrol. Earth Syst. Sci.}, 2\penalty0 (2/3):\penalty0 183--194,
  1998.

\bibitem[Fry and Swain(2010)]{fry2010hydrological}
M.~J. Fry and O.~Swain.
\newblock Hydrological data management systems within a national river flow
  archive.
\newblock Technical report, British Hydrological Society, 2010.

\bibitem[Furman(2008)]{furman2008modeling}
A.~Furman.
\newblock Modeling coupled surface--subsurface flow processes: A review.
\newblock \emph{Vadose Zone J.}, 7\penalty0 (2):\penalty0 741--756, 2008.

\bibitem[Gardner et~al.(1990)Gardner, Cooper, Wellings, Bell, and
  Hodnett]{gardner1990hydrology}
C.~M.~K. Gardner, J.~D. Cooper, S.~R. Wellings, J.~P. Bell, and M.~G. Hodnett.
\newblock Hydrology of the unsaturated zone of the chalk of south-east
  {England}.
\newblock In \emph{International chalk symposium}, pages 611--618, 1990.

\bibitem[Grieve et~al.(2016)Grieve, Mudd, and Hurst]{grieve2016long}
S.~W.~D. Grieve, S.~M. Mudd, and M.~D. Hurst.
\newblock How long is a hillslope?
\newblock \emph{Earth Surf. Processes Landforms}, 41\penalty0 (8):\penalty0
  1039--1054, 2016.

\bibitem[Henderson(1963)]{henderson1963stability}
F.~M. Henderson.
\newblock Stability of alluvial channels.
\newblock \emph{Trans. of the Amer. Soc. of Civ. Eng.}, 128\penalty0
  (1):\penalty0 657--686, 1963.

\bibitem[Horton(1932)]{horton1932drainage}
R.~E. Horton.
\newblock Drainage-basin characteristics.
\newblock \emph{Transactions, American geophysical union}, 13\penalty0
  (1):\penalty0 350--361, 1932.

\bibitem[Hutton et~al.(2016)Hutton, Wagener, Freer, Han, Duffy, and
  Arheimer]{hutton2016most}
C.~Hutton, T.~Wagener, J.~Freer, D.~Han, C.~Duffy, and B.~Arheimer.
\newblock Most computational hydrology is not reproducible, so is it really
  science?
\newblock \emph{Water Resour. Res.}, 52\penalty0 (10):\penalty0 7548--7555,
  2016.

\bibitem[Jones and Robins(1999)]{jones1999chalk}
H.~K. Jones and N.~S. Robins.
\newblock \emph{The chalk aquifer of the {South Downs}}.
\newblock British Geological Survey, 1999.

\bibitem[Kaufman and Rousseeuw(2009)]{kaufman2009finding}
L.~Kaufman and P.~J. Rousseeuw.
\newblock \emph{Finding groups in data: an introduction to cluster analysis}.
\newblock John Wiley \& Sons, 2009.

\bibitem[Kirkham(2014)]{kirkham2014principles}
M.~B. Kirkham.
\newblock \emph{Principles of soil and plant water relations}.
\newblock Academic Press, 2014.

\bibitem[Kjeldsen et~al.(2008)Kjeldsen, Jones, and
  Bayliss]{kjeldsen2008improving}
T.~R. Kjeldsen, D.~A. Jones, and A.~C. Bayliss.
\newblock \emph{Improving the {FEH} statistical procedures for flood frequency
  estimation}.
\newblock Environment Agency, 2008.

\bibitem[Kolditz et~al.(2012)Kolditz, Bauer, Bilke, B{\"o}ttcher, Delfs,
  Fischer, G{\"o}rke, Kalbacher, Kosakowski, McDermott,
  et~al.]{kolditz2012opengeosys}
O.~Kolditz, S.~Bauer, L.~Bilke, N.~B{\"o}ttcher, J.-O. Delfs, T.~Fischer, U.~J.
  G{\"o}rke, T.~Kalbacher, G.~Kosakowski, C.~I. McDermott, et~al.
\newblock Opengeosys: an open-source initiative for numerical simulation of
  thermo-hydro-mechanical/chemical ({THM/C}) processes in porous media.
\newblock \emph{Environ. Earth Sci.}, 67\penalty0 (2):\penalty0 589--599, 2012.

\bibitem[Kollet and Maxwell(2006)]{kollet2006integrated}
S.~J. Kollet and R.~M. Maxwell.
\newblock Integrated surface--groundwater flow modeling: {A} free-surface
  overland flow boundary condition in a parallel groundwater flow model.
\newblock \emph{Adv. Water Resour.}, 29\penalty0 (7):\penalty0 945--958, 2006.

\bibitem[Kristensen and Jensen(1975)]{kristensen1975model}
K.~J. Kristensen and S.~E. Jensen.
\newblock A model for estimating actual evapotranspiration from potential
  evapotranspiration.
\newblock \emph{Hydrol. Res.}, 6\penalty0 (3):\penalty0 170--188, 1975.

\bibitem[Lawley(2009)]{lawley2009soil}
R.~Lawley.
\newblock The soil-parent material database: a user guide.
\newblock Technical report, British Geological Survey, 2009.

\bibitem[Leibowitz et~al.(2018)Leibowitz, Wigington~Jr, Schofield, Alexander,
  Vanderhoof, and Golden]{leibowitz2018connectivity}
S.~G. Leibowitz, P.~J. Wigington~Jr, K.~A. Schofield, L.~C. Alexander, M.~K.
  Vanderhoof, and H.~E. Golden.
\newblock Connectivity of streams and wetlands to downstream waters: an
  integrated systems framework.
\newblock \emph{JAWRA J. Am. Water Resour. Assoc.}, 54\penalty0 (2):\penalty0
  298--322, 2018.

\bibitem[Lilley(2011)]{lilley2011ordnance}
B.~Lilley.
\newblock The ordnance survey {openData} initiative.
\newblock \emph{The Cartographic Journal}, 48\penalty0 (3):\penalty0 179--182,
  2011.

\bibitem[Liu et~al.(2004)Liu, Chen, Li, and Singh]{liu2004two}
Q.~Q. Liu, L.~Chen, J.~C. Li, and V.P. Singh.
\newblock Two-dimensional kinematic wave model of overland-flow.
\newblock \emph{J. Hydrol.}, 291\penalty0 (1-2):\penalty0 28--41, 2004.

\bibitem[Maxwell et~al.(2014)Maxwell, Putti, Meyerhoff, Delfs, Ferguson,
  Ivanov, Kim, Kolditz, Kollet, Kumar, et~al.]{maxwell2014surface}
R.~M. Maxwell, M.~Putti, S.~Meyerhoff, J.-O. Delfs, I.~M. Ferguson, V.~Ivanov,
  J.~Kim, O.~Kolditz, S.~J. Kollet, M.~Kumar, et~al.
\newblock Surface-subsurface model intercomparison: A first set of benchmark
  results to diagnose integrated hydrology and feedbacks.
\newblock \emph{Water Resour. Res.}, 50\penalty0 (2):\penalty0 1531--1549,
  2014.

\bibitem[Morawiecki(2022)]{github1}
P.~W. Morawiecki.
\newblock {GitHub} repository for parametric analysis of {UK} catchments.
\newblock \url{https://github.com/Piotr-Morawiecki/UK-catchment-properties},
  2022.
\newblock Accessed: 2022-10-26.

\bibitem[Mujumdar(2001)]{mujumdar2001flood}
P.~P. Mujumdar.
\newblock Flood wave propagation: The saint venant equations.
\newblock \emph{Resonance}, 6\penalty0 (5):\penalty0 66--73, 2001.

\bibitem[Nixon(1959)]{nixon1959study}
M.~Nixon.
\newblock A study of the bank-full discharges of rivers in {E}ngland and
  {W}ales.
\newblock \emph{Proc. Inst. Civ. Eng.}, 12\penalty0 (2):\penalty0 157--174,
  1959.

\bibitem[NRFA(2022)]{NRFA_dataset}
NRFA.
\newblock Catchment descriptors dataset.
\newblock
  \url{https://nrfaapps.ceh.ac.uk/nrfa/ws/station-info?station=*&format=html&fields=all},
  2022.
\newblock Accessed: 2022-10-26.

\bibitem[Price(1996)]{price1996introducing}
M.~Price.
\newblock \emph{Introducing groundwater}.
\newblock Taylor \& Francis, 1996.

\bibitem[Robins and Ball(2006)]{robins2006dumfries}
N.~Robins and D.~Ball.
\newblock \emph{The {Dumfries Basin} aquifer}.
\newblock British Geological Survey, 2006.

\bibitem[Robins and Buckley(1988)]{robins1988characteristics}
N.~S. Robins and D.~K. Buckley.
\newblock Characteristics of the {Permian} and {Triassic} aquifers of
  south-west {Scotland}.
\newblock \emph{Q. J. Eng. Geol. Hydrogeol.}, 21\penalty0 (4):\penalty0
  329--335, 1988.

\bibitem[Rodriguez-Iturbe and Rinaldo(2001)]{rodriguez2001fractal}
I.~Rodriguez-Iturbe and A.~Rinaldo.
\newblock \emph{Fractal river basins: chance and self-organization}.
\newblock Cambridge University Press, 2001.

\bibitem[Shaw et~al.(2010)Shaw, Beven, Chappell, and Lamb]{shaw2015hydrology}
E.~Shaw, K.~Beven, N.~Chappell, and R.~Lamb.
\newblock \emph{Hydrology in practice}.
\newblock CRC press, 3 edition, 2010.

\bibitem[Sitterson et~al.(2018)Sitterson, Knightes, Parmar, Wolfe, Avant, and
  Muche]{sitterson2018overview}
J.~Sitterson, C.~Knightes, R.~Parmar, K.~Wolfe, B.~Avant, and M.~Muche.
\newblock An overview of rainfall-runoff model types.
\newblock In \emph{Proceedings of 9th Int. Congr. Env. Mod. Soft}, 2018.

\bibitem[Song et~al.(2017)Song, Schmalz, Xu, and Fohrer]{song2017seasonality}
S.~Song, B.~Schmalz, Y.P. Xu, and N.~Fohrer.
\newblock Seasonality of roughness-the indicator of annual river flow
  resistance condition in a lowland catchment.
\newblock \emph{Water Resour. Manage.}, 31:\penalty0 3299--3312, 2017.

\bibitem[Stuetzle et~al.(2009)Stuetzle, Franklin, and
  Cutler]{stuetzle2009evaluating}
C.~Stuetzle, W.~R. Franklin, and B.~Cutler.
\newblock Evaluating hydrology preservation of simplified terrain
  representations.
\newblock \emph{SIGSPATIAL Special}, 1\penalty0 (1):\penalty0 51--56, 2009.

\bibitem[Tarboton and Ames(2001)]{tarboton2001advances}
D.~G. Tarboton and D.~P. Ames.
\newblock Advances in the mapping of flow networks from digital elevation data.
\newblock In \emph{Bridging the Gap: Meeting the World's Water and
  Environmental Resources Challenges}, pages 1--10, 2001.

\bibitem[Tayfur and Kavvas(1994)]{tayfur1994spatially}
G.~Tayfur and M.~L. Kavvas.
\newblock Spatially averaged conservation equations for interacting
  rill-interrill area overland flows.
\newblock \emph{J. Hydraul. Eng.}, 120\penalty0 (12):\penalty0 1426--1448,
  1994.

\bibitem[T{\'o}th et~al.(2017)T{\'o}th, Weynants, P{\'a}sztor, and
  Hengl]{toth20173d}
B.~T{\'o}th, M.~Weynants, L.~P{\'a}sztor, and T.~Hengl.
\newblock 3{D} soil hydraulic database of {E}urope at 250 m resolution.
\newblock \emph{Hydrol. Processes}, 31\penalty0 (14):\penalty0 2662--2666,
  2017.

\bibitem[Van~Genuchten(1980)]{van1980closed}
M.~Th. Van~Genuchten.
\newblock A closed-form equation for predicting the hydraulic conductivity of
  unsaturated soils.
\newblock \emph{Soil Sci. Soc. Am. J.}, 44\penalty0 (5):\penalty0 892--898,
  1980.

\bibitem[Vieira(1983)]{vieira1983conditions}
J.~H.~Daluz Vieira.
\newblock Conditions governing the use of approximations for the {Saint-Venant}
  equations for shallow surface water flow.
\newblock \emph{J. Hydrol.}, 60\penalty0 (1-4):\penalty0 43--58, 1983.

\bibitem[Ward et~al.(1997)Ward, Williams, and Chadha]{ward1997use}
R.~S. Ward, A.~T. Williams, and D.~S. Chadha.
\newblock The use of groundwater tracers for assessment of protection zones
  around water supply boreholes--a case study.
\newblock In \emph{Proceedings of the 7th International Symposium in water
  tracing, Portoroz, Slovenia}, page 369–375, 1997.

\bibitem[Waters et~al.(2016)Waters, Terrington, Cooper, Raine, and
  Thorpe]{waters2016construction}
C.~N. Waters, R.~L. Terrington, M.~R. Cooper, R.~J. Raine, and S.~Thorpe.
\newblock The construction of a bedrock geology model for the uk:
  {UK3D}\_v2015.
\newblock Technical report, British Geological Survey, 2016.

\bibitem[Weill et~al.(2009)Weill, Mouche, and Patin]{weill2009generalized}
S.~Weill, E.~Mouche, and J.~Patin.
\newblock A generalized {R}ichards equation for surface/subsurface flow
  modelling.
\newblock \emph{J. Hydrol.}, 366\penalty0 (1-4):\penalty0 9--20, 2009.

\bibitem[Wharton(1992)]{wharton1992flood}
G.~Wharton.
\newblock Flood estimation from channel size: guidelines for using the
  channel-geometry method.
\newblock \emph{Appl. Geogr.}, 12\penalty0 (4):\penalty0 339--359, 1992.

\bibitem[Yan and Smith(1994)]{yan1994simulation}
J.~Yan and K.~R. Smith.
\newblock Simulation of integrated surface water and ground water systems -
  model formulation.
\newblock \emph{{JAWRA} J. Am. Water Resour. Assoc.}, 30\penalty0 (5):\penalty0
  879--890, 1994.

\end{thebibliography}

\newpage
\appendix

\section{Data extraction methods implementation}

\noindent In this appendix, we describe the numerical methods used for data extraction and processing. In some cases, our methods require significant optimisation to reduce the processing time for large datasets. Only those parameters that require special handling are discussed here, namely $L_x^\mathrm{stream}$, $L_y^\mathrm{main}$ and $L_y^\mathrm{all}$, $L_z$, $S_x$ and $S_y$. Processing methods for other parameters were sufficiently covered in \cref{sec:typical_parameters}.

All data processing methods were implemented in the R programming language. The source code with further explanation of implementation details is publicly available in our GitHub repository \citep{github1}. 

\subsection{Extracting catchment length (\texorpdfstring{$L_y$}{})}
\label{app:extracting_catchment_length}

\noindent As described in \cref{sec:catchment_size}, in order to determine the length of rivers for each catchment, we use two alternative datasets: OS Open Rivers and OS VectorMap District. The first dataset is available in a single shape file for the entire UK. In this case, we find the intersection of all streams defined in the file with the boundary of each respective catchment. Then, we calculate the total length of the streams located inside the catchment.

\begin{figure}
    \centering
    \includegraphics{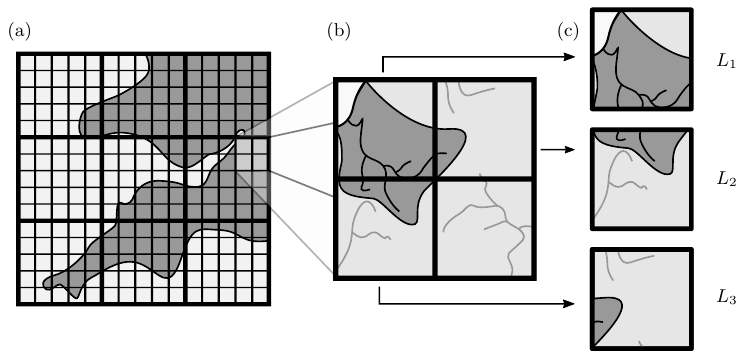}
    \caption{Illustration showing the division of the UK into tiles. The left-most map shows 9 National Grid tiles, each divided into 25 subtiles. \edit{The dark gray area in (b) represents the area from which the streamlines reach a given outlet (typically following the direction of steepest descent). The resultant catchment has an intersection with three tiles shown in (c).} When analysing each of these subtiles, all streams within a given subtile are overlapped with the catchment boundary, as illustrated in (c). Adding the lengths of these streams $L_1$, $L_2$ and $L_3$, allows us to calculate the total stream length for the given catchment.}
    \label{fig:subtiles}
\end{figure}

\noindent The OS VectorMap District dataset is divided into 55 Ordnance Survey National Grid reference system tiles of size $100~\mathrm{km}\times100~\mathrm{km}$. Since the amount of spatial data in each of these tiles is large, we further divide them into $5\times 5$ subtiles. For each subtile, we find all rivers (and other surface water bodies) located within its boundary, as illustrated in \cref{fig:subtiles}. Then, we find the intersection of this river subset with each catchment, which is overlapping the given subtile. By introducing subtiles, we significantly reduce the computational time. Determining whether a given river is within the square subtile is relatively fast compared to finding the intersection of a river with an irregular catchment boundary. This approach allows us to significantly limit the number of rivers that need to be intersected with the boundary of each catchment. As described in \cref{sec:catchment_size}, for surface water bodies represented using polygons (usually corresponding to wide rivers), the length is defined as half the polygon's perimeter. We repeat this procedure for all National Grid tiles and their subtiles, summing the length of all streams intersecting with each catchment.

\subsection{Extracting hillslope width (\texorpdfstring{$L_x$}{}) and gradient (\texorpdfstring{$S_x$}{})}
\label{app:extracting_hillslope_width_gradient}

The OS Terrain 50 dataset includes a Digital Terrain Model (DTM) with a resolution of $50$~m. The dataset is represented as a raster divided into Ordnance Survey National Grid tiles and further subdivided into $10\times 10$ subtiles (size of $10~\mathrm{km}\times10~\mathrm{km}$).

\begin{figure}
    \centering
    \includegraphics{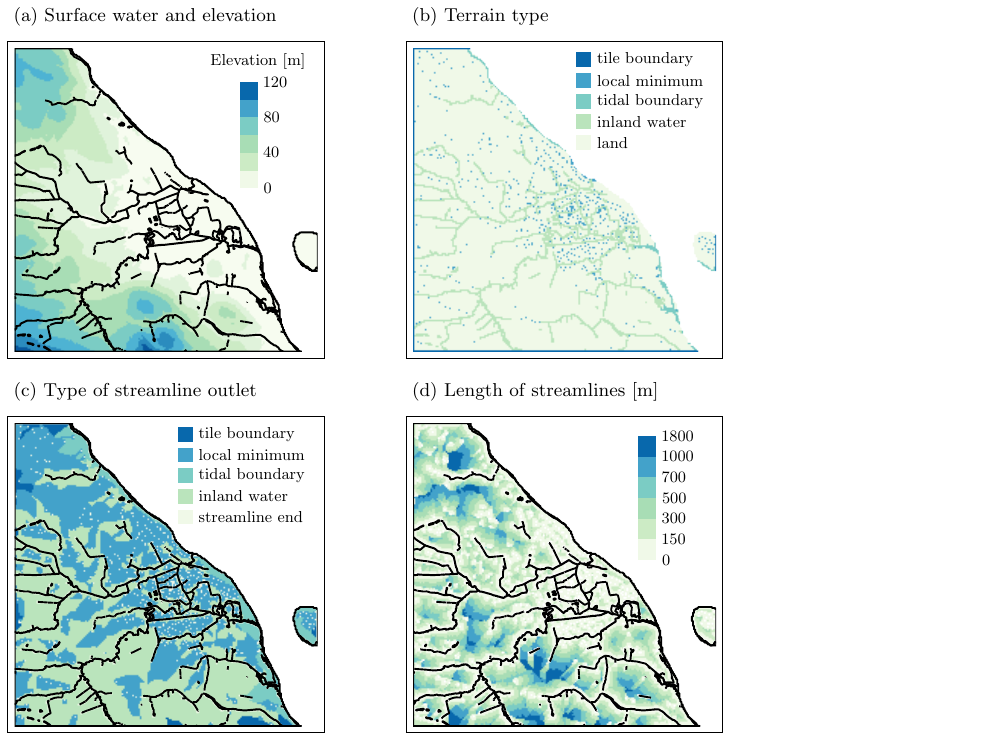}
    \caption{Streamline analysis steps. Fig. (a) presents the input datasets: digital terrain model (DTM) in raster form with surface water bodies (black lines and polygons). These datasets are used to construct the terrain-type raster presented in fig. (b). Then, from each land tile a streamline is generated until reaching another terrain type, which is presented in fig. (c). For each starting point, the length of each streamline is presented in fig. (d).}
    \label{fig:streamline_processing_steps}
\end{figure}

To extract the hillslope width and gradient, we create an additional raster for each subtile that stores information about the type of terrain in the given tile. The types are as follows: 1 -- tidal boundary (\emph{e.g.} coast), 2 -- large channel/lake (represented as a polygon), 3 -- small channel (represented as a line), 4 -- subtile boundary, and 5 -- local elevation minimum (without surface water). Types 1--3 are determined by overlapping different surface water bodies from the OS VectorMap dataset District with a raster representing the given subtile. We determine a local elevation minimum based on the altitude of the neighbouring raster elements. An example of a terrain-type raster is presented in \cref{fig:streamline_processing_steps}.

For all raster tiles that were not classified into one of the above categories, we find one out of the eight neighbouring tiles that is located in the direction of the steepest descent. Then, for each element, we iteratively construct the steepest descent path, until it reaches one of the specified terrain types. This procedure allows us to find overland streamlines that follow the hillslope gradient. When the path is found for all raster tiles, we save their length and the elevation difference between the start and end point. Then, for each catchment, we aggregate the values assigned to all raster tiles located inside it and ending in rivers (terrain types 2 or 3). The streamlines ending in the tidal boundary, local elevation minima, or leaving through the subtile boundary, are not taken into account.

To find the catchment width, we calculate the mean streamflow path and multiply it by 2, as discussed in \cref{sec:estimating_Lx}. To find the elevation gradient along the hillslope, we divide the total elevation difference along all paths by the total path length, as discussed in \cref{sec:estimating_Sx}.

\subsection{Extracting gradient along a river (\texorpdfstring{$S_y$}{})}
\label{app:extracting_gradient_along_river}

\noindent To estimate the typical value of the gradient along the rivers, we use the OS Terrain 50 and OS Open Rivers datasets. For each subtile represented in the OS Terrain dataset, we extract rivers that overlap with the given subtile. Then, for each node of the river, we find its elevation. The mean gradient $S_y$ is estimated as the total elevation difference along the river segments divided by their total length, as discussed in \cref{sec:estimating_Sy}. As in the case of the $S_x$ estimation method, this measure is not highly sensitive to local terrain drops along the river.

\subsection{Extracting aquifer thickness (\texorpdfstring{$L_z$}{})}
\label{app:extracting_aquifer_thickness}

\begin{figure}
    \centering
    \includegraphics[trim={0 0 2.5cm 0},clip]{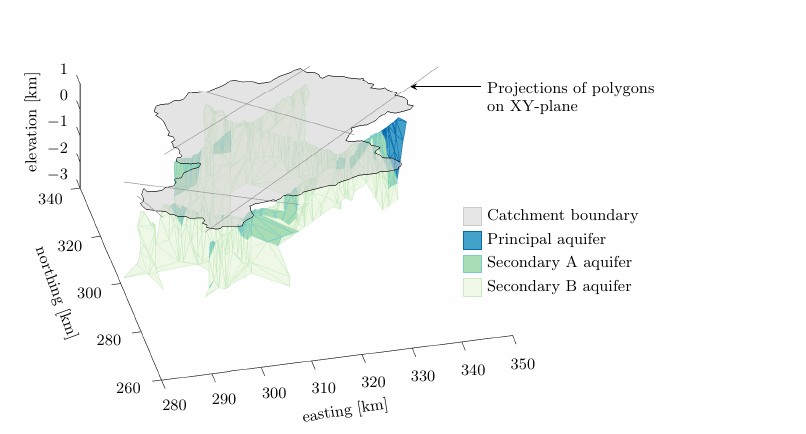}
    \caption{A fence diagram showing the hydrological properties of the aquifer of the River Avon catchment with the outlet at Bath St James. Data was extracted from the UK3D dataset. The fence diagram for the entire UK can be generated using the Geology of Britain viewer~\citep{GB_viewer}.}
    \label{fig:fence_diagram}
\end{figure}

Extracting aquifer thickness at the catchment scale is a challenging task, since the Bedrock Geology Model by the British Geological Society represents it in the form of three-dimensional polygons located sparsely across the UK, and the data is not appropriately aligned in the $xy$-plane. A sample of this dataset is presented in \cref{fig:fence_diagram}. The idea of extracting the mean aquifer thickness is to calculate the total area of polygons within a given catchment extent, and then divide it by the estimated length of these cross-sections projected onto the $xy$-plane.

\begin{figure}
    \centering
    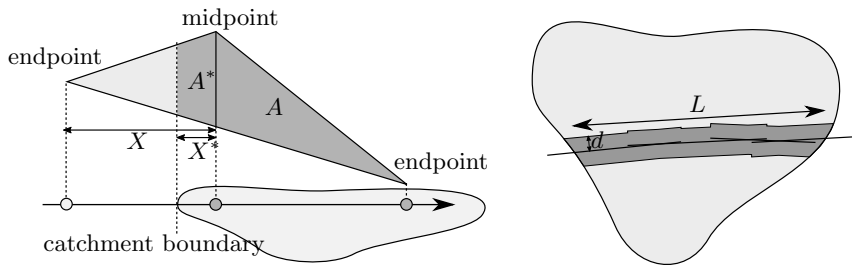
    \caption{On the left: Method of dividing triangular polygons in order to calculate their area within the catchment. On the right: Method of estimating the cross-section length $L$ inside the catchment.}
    \label{fig:dividing_polygons}
\end{figure}

To find the total aquifer area, we divide each polygon into triangular parts. Then, each triangle is further divided into two triangular parts with a common edge directed along the $z$-axis, as illustrated in \cref{fig:dividing_polygons}. We refer to the point on the $xy$-plane that corresponds to the common edge as the 'midpoint', while the other two points are referred to as 'endpoints'. For each catchment, we find all triangular polygons, with at least one of these three points located inside the given catchment.

We need to find the area of each polygon within the catchment boundary. If both endpoints are inside the catchment, we take the total polygon area $A$. If any endpoint is outside the catchment, we calculate the area of the part of this polygon that is located inside the catchment. This area, denoted as $A^*$, is expressed either as $A^*=\left(\frac{x^*}{x}\right)^2A$ if the midpoint is outside or $A^*=A-\left(\frac{x^*}{x}\right)^2A$ if the endpoint is inside the catchment. Here, $x^*$ denotes the length of the triangle projection in the xy-plane included inside the catchment boundary (see \cref{fig:dividing_polygons}). The area of all triangular polygons within the catchment is then summed up.

Finding the projection of the profiles on the catchment area is challenging since the triangular polygons in the BGS database are not aligned, and in addition, there are many polygons. To solve this problem, we first convert each polygon projection into a thin rectangular polygon of width $d$, with its centreline going along the given projection. Then, we find the union of all polygons intersecting a given catchment (see \cref{fig:dividing_polygons}). The length of the cross-section can be estimated in two ways: either as the area of the union divided by $2d$, or as the perimeter of the union. In both cases, there is a small overestimation of the true length $L$, which is minimised by choosing a value of $d$ that is significantly larger than the misalignment of polygon projections, and at the same time significantly smaller than typical catchment dimensions. In this study, we estimated the cross-section length by calculating the union's area obtained for rectangular polygons with a width $d=15$~m. Dividing the total area by the total length gives us the mean thickness of the aquifer.

\section{Spatial distribution of catchment parameters}
\label{app:spatial_distribution}

\noindent
In this study, we estimated the physical parameters for 1601 catchments in the UK as defined in the National River Flow Archive (NRFA). The resultant database is available in the project's GitHub repository. A sample of the constructed dataset, including the parameters for the first five catchments, is presented in \cref{tab:catchment_parameters_sample}.

\begin{table}
    \centering
    \begin{tabular}{lccccc}
        id & 1001 & 2001 & 2002 & 3001 & 3002 \\
        area & 1.62e+08 & 5.51e+08 & 4.34e+08 & 4.95e+08 & 2.41e+08 \\
        rainfall & 2.96e-08 & 3.54e-08 & 3.86e-08 & 5.07e-08 & 5.65e-08 \\
        runoff & 1.86e-08 & 2.40e-08 & 2.76e-08 & 3.07e-08 & 3.76e-08 \\
        evapotranspiration & 1.10e-08 & 1.14e-08 & 1.10e-08 & 2.00e-08 & 1.89e-08 \\
        Manning\_n\_hillslope & 0.0452 & 0.0363 & 0.0353 & 0.0560 & 0.0377 \\
        channel\_width & 1.58 & 3.32 & 3.15 & 3.55 & 2.74 \\
        channel\_depth & 0.787 & 1.29 & 1.25 & 1.35 & 1.14 \\
        width\_B & 687. & 725. & 764. & 709. & 759. \\
        gradient\_perpendicular & 0.0264 & 0.0982 & 0.10166 & 0.117 & 0.208 \\
        length\_A & 1.66e+05 & 3.01e+05 & 2.57e+05 & 3.17e+05 & 1.58e+05 \\
        gradient\_parallel & 0.0129 & 0.0183 & 0.0225 & 0.0259 & 0.0520 \\
        length\_B & 5.13e+05 & 9.31e+05 & 6.60e+05 & 9.88e+05 & 4.80e+05 \\
        aquifer\_thickness & 699. & 0 & 0 & 0 & 0 \\
        hydraulic\_conductivity\_A & 3.20e-06 & 3.20e-06 & 3.20e-06 & NA & 3.20e-06 \\
        MvG\_alpha & 5.42 & 7.19 & 7.37 & NA & 7.45 \\
        MvG\_thetaR & 0 & 0 & 0 & NA & 0 \\
        MvG\_thetaS & 0.467 & 0.444 & 0.442 & NA & 0.441 \\
        MvG\_n & 1.1918 & 1.1906 & 1.1905 & NA & 1.1904 \\
        width\_A & 315. & 592. & 658. & 500. & 501. 
    \end{tabular}
    \caption{A sample of dataset consisting of estimates of physical parameters for all UK catchments included in NRFA. The first column represents the names of the parameters used in our dataset. The top row represents the ID numbers assigned to the given catchments in the NRFA database. The detailed description of all catchments can be found in the~\cite{NRFA_dataset} dataset.} 
    \label{tab:catchment_parameters_sample}
\end{table}

\begin{landscape}
    \begin{figure}
        \includegraphics[angle=90]{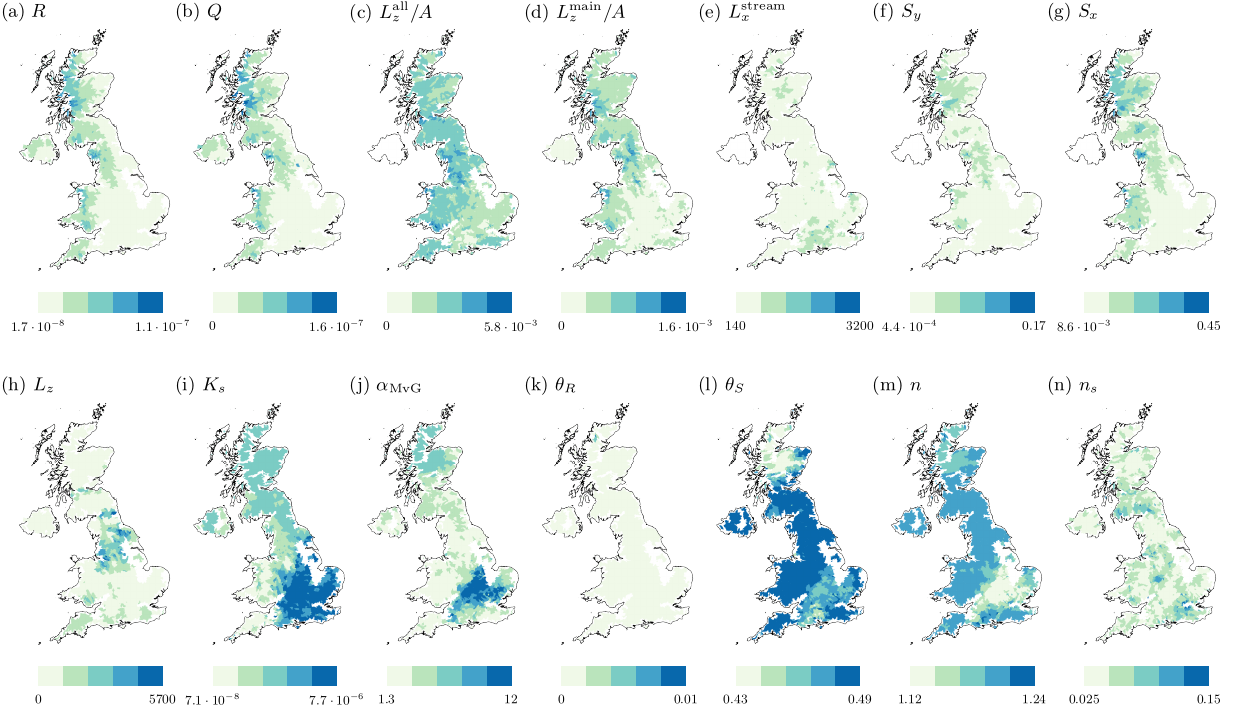}
        \caption{Map of UK catchments listed in NRFA with extracted mean values of the physical parameters.}
        \label{fig:UK_map_grid}
    \end{figure}
\end{landscape}

In \cref{sec:PCA}, we discuss the results of applying principal component analysis to the catchment dataset. Each principal component is defined as a linear combination of the selected physical parameters from \cref{tab:catchment_parameters_sample}:
$$PC_k=\sum_i=1^n w_{i,k} x_i$$
\noindent where $x_i$ is the $i$\textsuperscript{th} physical parameter, and $w_{i,k}$ is the weight of the $i$\textsuperscript{th} parameter and $k$\textsuperscript{th} principal component. The weights for the first five principal components are presented in \cref{tab:PCA_table}. Additionally, the spatial distribution of the values of the selected catchment parameters from \cref{tab:catchment_parameters_sample} is presented in \cref{fig:UK_map_grid}.

\begin{table}
    \centering
    \begin{tabular}{rS[table-format=3.4]S[table-format=3.4]S[table-format=3.4]S[table-format=3.4]S[table-format=3.4]c}
        & \multicolumn{1}{c}{PC1} & \multicolumn{1}{c}{PC2} & \multicolumn{1}{c}{PC3} & \multicolumn{1}{c}{PC4} & \multicolumn{1}{c}{PC5} & ... \\
        $R$ & 0.405 & -0.227 & 0.064 & -0.079 & -0.152 & \\
        $Q$ & 0.395 & -0.214 & 0.043 & -0.118 & -0.147 & \\
        $n_s$ & -0.011 & 0.255 & 0.333 & -0.822 & -0.258 & \\
        $L_x^\mathrm{stream}$ & -0.212 & -0.124 & 0.672 & 0.323 & -0.157 & \\
        $S_x$ & 0.391 & -0.259 & 0.042 & -0.041 & -0.032 & \\
        $S_y$ & 0.390 & -0.217 & 0.001 & -0.005 & -0.171 & \\
        $L_z$ & 0.038 & 0.417 & -0.117 & 0.312 & -0.821 & \\
        $K_s$ & -0.363 & -0.088 & 0.213 & -0.203 & -0.055 & \\
        $\alpha_\mathrm{MvG}$ & -0.290 & -0.443 & -0.276 & -0.115 & -0.263 & \\
        $\theta_s$ & 0.138 & 0.481 & -0.326 & -0.124 & 0.166 & \\
        $n$ & 0.306 & 0.314 & 0.434 & 0.175 & 0.244 &
    \end{tabular}
    \caption{The first five principal components obtained from UK catchments.}
    \label{tab:PCA_table}
\end{table}

\end{document}